\documentclass[reprint,sort&compress,superscriptaddress,amsmath,aps,showpacs,pra]{revtex4-1}
\usepackage{amssymb}
\usepackage{amsmath}
\usepackage{bbold}
\usepackage{appendix}
\usepackage{verbatim}
\usepackage{float}
\usepackage{color}
\usepackage{textcomp}
\usepackage{gensymb}
\usepackage[vietnamese, english]{babel}
\usepackage{hyperref}
\hypersetup{
     colorlinks   = true,
     citecolor    = blue
}
\usepackage{siunitx}
\usepackage{graphicx}
\usepackage{dcolumn}
\usepackage{bm}
\usepackage{braket}
\usepackage{empheq}
\usepackage{nicefrac} 
\usepackage[dvipsnames]{xcolor}
\renewcommand{\vec}[1]{\bm{#1}}

\newcommand{\be}{\begin{equation}}
\newcommand{\ee}{\end{equation}}
\newcommand{\bea}{\begin{eqnarray}}
\newcommand{\eea}{\end{eqnarray}}

\def\nn{\nonumber}

\def\pref{\eqref}

\begin{document}

\title{Band Structure and Superconductivity in Twisted Trilayer Graphene}

\author{\foreignlanguage{vietnamese}{Võ Tiến Phong}}
\affiliation{Department of Physics and Astronomy, University of Pennsylvania, Philadelphia PA 19104}
\author{Pierre A. Pantale\'on}
\affiliation{Imdea Nanoscience, Faraday 9, 28015 Madrid, Spain}
\author{Tommaso Cea}
\affiliation{Imdea Nanoscience, Faraday 9, 28015 Madrid, Spain}
\author{Francisco Guinea}
\affiliation{Imdea Nanoscience, Faraday 9, 28015 Madrid, Spain}
\affiliation{Donostia International Physics Center, Paseo Manuel de Lardiz\'abal 4, 20018 San Sebasti\'an, Spain}

\date{\today}

\begin{abstract}
We study the symmetries of twisted trilayer graphene's band structure under various extrinsic perturbations, and analyze the role of long-range electron-electron interactions near the first magic angle. The electronic structure is modified by these interactions in a similar way to twisted bilayer graphene. We analyze electron pairing due to long-wavelength charge fluctuations, which are coupled among themselves via the Coulomb interaction and additionally mediated by longitudinal acoustic phonons. We find superconducting phases with either spin singlet/valley triplet or spin triplet/valley singlet symmetry, with critical temperatures of up to a few Kelvin for realistic choices of parameters.

\end{abstract}

\maketitle

\textit{Introduction:} Recently, it has been shown that superconductivity and exotic correlated phases can emerge when two or three monolayers of graphene are laterally stacked with a small relative twist angle between successive layers~\cite{Cao2018,Cao2018_bis,Park2021,Hao2021}. Observation of insulating behavior and/or superconductivity have also been reported for graphene bilayers on hBN substrates~\cite{Moriyama2019,Shen2019}, ABC-stacked trilayers on hBN substrates~\cite{Chen2019,Chen2019a,Chen2020, Chittari2019a}, pairs of graphene bilayers twisted with respect to each other~\cite{HeSym2020,Letal19,Cetal19,Tsai2019}, rhombohedral tetralayers~ \cite{Kerelsky2019b}, rhombohedral trilayers~\citep{zhou2021,zhou2021a} and twisted transition-metal-dichalcogenide layers~\cite{Wang2019b}. However, twisted bilayer graphene (TBG)~\cite{Cao2018,Cao2018_bis} and its recent cousin, the alternating-angle twisted trilayer graphene (TTG)~\citep{Park2021,Hao2021}, are the only systems to date where superconductivity has been unambiguously established~\cite{Lu2019,Arora2020,Yankowitz2019}. 

Twisted trilayer graphene with alternating angles shows a series of magic angles with remarkably flat bands, where, under certain conditions, the trilayer system behaves as a TBG with renormalized interlayer hoppings plus a monolayer~\cite{KKTV19}. This follows from the mirror symmetry around the central layer of the system. This symmetry can be broken by various experimentally-relevant perturbations. In this case,  calculations of the electronic structure in the absence of this mirror symmetry shows hybridized monolayer and twisted bilayer features~\cite{Park2021,WZY21,RL21,CXSLRB21,SCJ21,FGMLKK21}. The effect of electron-electron interactions have also been considered~\cite{RL21,CXSLRB21,SCJ21,FGMLKK21,LS21,Getal21,CSS21}, as well as the role of external magnetic fields~\cite{QM21}. 

In this work, we study the symmetries of the continuum model, the effects of  symmetry-breaking perturbations on the band structure, and the renormalization of electronic energies by electron-electron interactions within a self-consistent Hartree formalism, complemented by the analysis of the exchange potential at half filling. Then by using  an extended Kohn-Luttinger mechanism that includes electron-hole pairs, plasmons, and phonons, we  analyze how the screened Coulomb interaction can induce pairing in TTG.

\begin{figure*}
\includegraphics[scale=0.55]{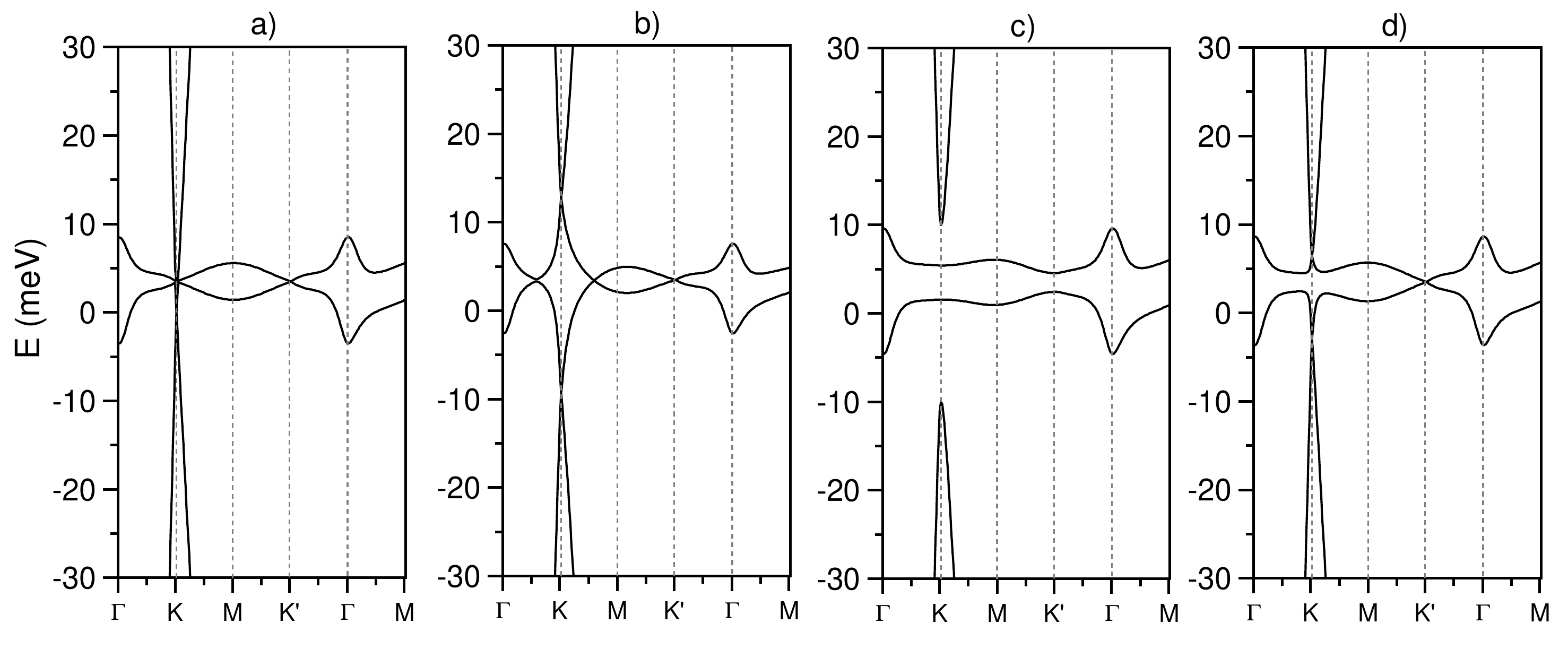}
\caption{\textbf{Single-particle band structure with various perturbations:} (a) Intrinsic TTG with a twist angle of $\theta = 1.59^\circ$ where the monolayer band is decoupled from the narrow bilayer bands. (b) A perpendicular displacement field $\Delta V = 50 $~meV between the outermost layers hybridizes these bands. (c) A non-zero staggered sublattice potential, $\delta_{1} = \delta_{3} = 10$ meV, generically breaks both $\mathcal{T}C_{2z}$ and $M_z$ and gaps out the Dirac cones near charge neutrality, endowing the resulting bands with possible nonzero Chern numbers. (d) For the  special case, $\delta_{1} = -\delta_{3},$ while both $\mathcal{T}C_{2z}$ and $M_z$ are broken individually, $\mathcal{T}C_{2z}M_z$ is preserved. In this case, we find that the bands are mixed, but the Dirac cones are preserved, though they can be pushed to different energies.}
\label{fig: Bands}
\end{figure*}

\textit{Band structure and symmetries of the continuum model:} To form TTG, we consider three monolayers stacked one on top of the other in perfect atomic registry, i.e. in the $AA$ stacking configuration. Labeling the layers consecutively, we twist layers $\ell = 1$ and $\ell = 3$ by $-\theta/2$ and layer $\ell = 2$ by $+\theta/2$ about a fixed hexagon center. To describe the low-energy physics at small angles, we adopt a valley-projected continuum Hamiltonian~\cite{BM11, KYK18,KKTV19, LWM19}
\begin{equation}
\begin{split}
\mathcal{H}^{\theta}_{\nu} (\mathbf{k}, \mathbf{r}) &= \begin{pmatrix}
\mathfrak{h}_\nu \left( \mathbf{k}^{+\frac{\theta}{2}} \right) +V_1 &  \left[\mathfrak{t}_\nu^\theta (\mathbf{r}) \right]^\dagger & \mathfrak{t}_\text{AA}\\
\mathfrak{t}_\nu^\theta (\mathbf{r}) &\mathfrak{h}_\nu \left( \mathbf{k}^{-\frac{\theta}{2}} \right) &  \mathfrak{t}_\nu^\theta (\mathbf{r}) \\
\mathfrak{t}_\text{AA}^\dagger & \left[ \mathfrak{t}_\nu^\theta (\mathbf{r}) \right]^\dagger & \mathfrak{h} _\nu\left( \mathbf{k}^{+\frac{\theta}{2}}\right) + V_3
\end{pmatrix}, \\
\mathfrak{h}_\nu \left( \mathbf{k} \right) &= - \hbar v_F \left( \mathbf{k} - \mathbf{K}_\nu \right) \left( \nu \sigma_x, \sigma_y \right), \\
\mathfrak{t}_\nu^\theta (\mathbf{r})  &= \begin{pmatrix}
w_0 & w_1 \\
w_1 & w_0
\end{pmatrix} + \begin{pmatrix}
w_0 & w_1 e^{-i \phi} \\
w_1 e^{i \phi}& w_0
\end{pmatrix} e^{i \nu \mathbf{G}_1^\text{M} \cdot \mathbf{r}}\\
&  + \begin{pmatrix}
w_0 & w_1 e^{i \phi} \\
w_1 e^{-i \phi}& w_0
\end{pmatrix} e^{i \nu \left(\mathbf{G}_1^\text{M}+\mathbf{G}_2^\text{M} \right) \cdot \mathbf{r}}, \\
\mathfrak{t}_\text{AA} &= \begin{pmatrix}
\gamma_2 & 0 \\
0 & \gamma_2
\end{pmatrix}
\end{split}
\label{eq: Hamiltonian}
\end{equation}
where $\phi = {2 \pi \nu /3}$, $\hbar v_F/a = 2.135 $ eV, $w_0 = 79.7$ meV and $w_1 = 97.5$ meV are the interlayer tunneling amplitudes whose imbalance accounts for some lattice relaxation~\cite{KYK18}, $\nu = \pm$ is the valley index, $\mathbf{k}^{\theta} = R(\theta) \mathbf{k}$,  $\mathbf{K}_\pm = \frac{4 \pi}{3a} \left( \mp 1, 0 \right)$ are the microscopic unrotated zone corners, and $\gamma_2$ is a small $AA$ coupling between the first and third layers.  If we use the conventional Slater-Koster parametrization~\cite{KYK18}, $\gamma_2 \approx 0.3$ meV, which justifies its neglect hereafter. It is included here to clarify some symmetry properties to be discussed. We note in passing that this approximation warrants further inspection in future work since it plays an important role in the modeling of graphite's band structure \cite{M57, SW58}. To study various perturbations~\citep{Cea2020TBGhBN,Shi2021tBGhBN,Shin2021tBGhBN,Mao2021tBGhBN,Lin2021tBGhBN}, we also add on-site couplings to the first and third layers. For substrate induced masses, $V_{1}=\delta_1 \sigma_z, V_3 = \delta_3 \sigma_z,$ while for a perpendicular displacement field, $V_{1}, V_3 \propto \mathbb{1}.$ Direct coupling to $\ell = 2$ is ignored since this layer is encapsulated and therefore difficult to access.

We now consider the symmetries of Hamiltonian \eqref{eq: Hamiltonian} with $V_{1} = V_3 = 0$.  Our choice of twist center ensures that the structure is maximally symmetric. A different twist center, for instance, about a registered site, retains only a subgroup of the maximal group. The vertical rotations about the $z$ axis, $C_{6z},$ are respected by each plane individually. The horizontal rotations about the in-plane axes, $C_{2x}$ and $C_{2y},$ preserved by TBG, are explicitly broken due to the presence of the middle layer in TTG. However, reflection symmetry about the plane of the middle layer $M_z$ that exchange the top and bottom layers is preserved in TTG. Alternatively, one can substitute $M_z$ for inversion symmetry $\mathcal{I}$ that is equivalent to $M_z$ followed by a rotation.  Therefore, the point symmetry group of the lattice is $C_{6h} \otimes \mathcal{T},$ where $\mathcal{T}$ is local time-reversal symmetry. In addition to these microscopic point symmetries, we can also now impose valley projection symmetry which becomes exact in the limit of large moir\'{e} wavelength. In this case, only a subset of  $C_{6h} \otimes \mathcal{T}$ is preserved at a single valley, namely $C_{2z}$ is broken because it takes $\mathbf{K}_+$ to $\mathbf{K}_-$ in momentum space. The symmetry group of the Hamiltonian thus factorizes into a direct product of valley symmetry and a magnetic point symmetry group that consists of $\lbrace C_{3z}, \mathcal{T} C_{2z}, M_z \rbrace.$ 

The presence of $z$-mirror symmetry allows us to simplify Eq. \eqref{eq: Hamiltonian} significantly. Mirror symmetry leaves the middle layer, $\ell = 2,$ invariant, while it interchanges layers $\ell = 1$ and $\ell = 3.$ So we can form linear combinations of the wavefunctions of these two exchanged layers that are odd and even under mirror symmetry. Such a transformation is executed by the unitary operator $\mathcal{U}$~\cite{LWM19, KKTV19}
\begin{equation}
\begin{pmatrix}
\ket{\mathcal{M}_\text{o}^1} \\
\ket{\mathcal{M}_\text{o}^2} \\
\ket{\mathcal{M}_\text{e}} \\
\end{pmatrix} = \frac{1}{\sqrt{2}}\begin{pmatrix}
1 & 0 & 1  \\
0 & \sqrt{2} & 0 \\
-1 & 0 & 1 \\
\end{pmatrix}\begin{pmatrix}
\ket{\ell = 1} \\
\ket{\ell = 2} \\
\ket{\ell = 3} \\
\end{pmatrix}.
\end{equation}
Under this transformation, the Hamiltonian factorizes into 
\begin{equation}
\begin{split}
&\tilde{\mathcal{H}}_\nu^\theta(\mathbf{k}, \mathbf{r}) = \mathcal{U}\mathcal{H}_\nu^\theta(\mathbf{k}, \mathbf{r}) \mathcal{U}^\dagger = \tilde{\mathcal{H}}_\text{TBG}^{\theta, \nu} \left( \mathbf{k}, \mathbf{r} \right) \oplus \tilde{\mathcal{H}}_\text{MLG}^{\theta, \nu} \left( \mathbf{k} \right) \\
&= \begin{pmatrix}
\mathfrak{h}_\nu \left( \mathbf{k}^{+\frac{\theta}{2}} \right) + \mathfrak{t}_\text{AA}  &  \sqrt{2}\left[\mathfrak{t}_\nu^\theta (\mathbf{r}) \right]^\dagger & 0\\
\sqrt{2}\mathfrak{t}_\nu^\theta (\mathbf{r}) &\mathfrak{h}_\nu \left( \mathbf{k}^{-\frac{\theta}{2}} \right) &  0 \\
0 & 0 & \mathfrak{h} _\nu\left( \mathbf{k}^{+\frac{\theta}{2}}\right) -\mathfrak{t}_\text{AA} 
\end{pmatrix}, 
\end{split}
\label{eq: HamiltonianNewBasis}
\end{equation}
where the mirror-odd sector, $ \tilde{\mathcal{H}}_\text{TBG}^{\theta, \nu} \left( \mathbf{k}, \mathbf{r} \right),$ behaves as though it were a TBG system with enhanced interlayer couplings by a factor of $\sqrt{2}$ and the mirror-even sector, $ \tilde{\mathcal{H}}_\text{MLG}^{\theta, \nu} \left( \mathbf{k}, \mathbf{r} \right),$ mimics a single rotated layer of graphene. This transformation is convenient because it casts the Hamiltonian into a direct sum of two Hamiltonians that are well-understood. Denoting operators in the new basis with an overline tilde, we see that $\tilde{C}_{3z} = C_{3z}$ and $\tilde{\mathcal{T}}\tilde{C}_{2z} = \mathcal{T} C_{2z}$ since these preserve layer index. In the mirror-odd sector, there are additional symmetries thus far neglected that exchanges basis states $\ket{\mathcal{M}_\text{o}^1}$ and $\ket{\mathcal{M}_\text{o}^2}$ when $\gamma_2 = 0.$ We can call them $\tilde{C}_{2x}$ and $ \mathcal{T}\tilde{C}_{2y},$ although they do not correspond naturally to rotations in the original layer basis. When $\gamma_2 \neq 0,$ these symmetries are broken, but all the other symmetries of the lattice in the layer basis are still preserved. 

\begin{figure*}
\includegraphics[scale=0.55]{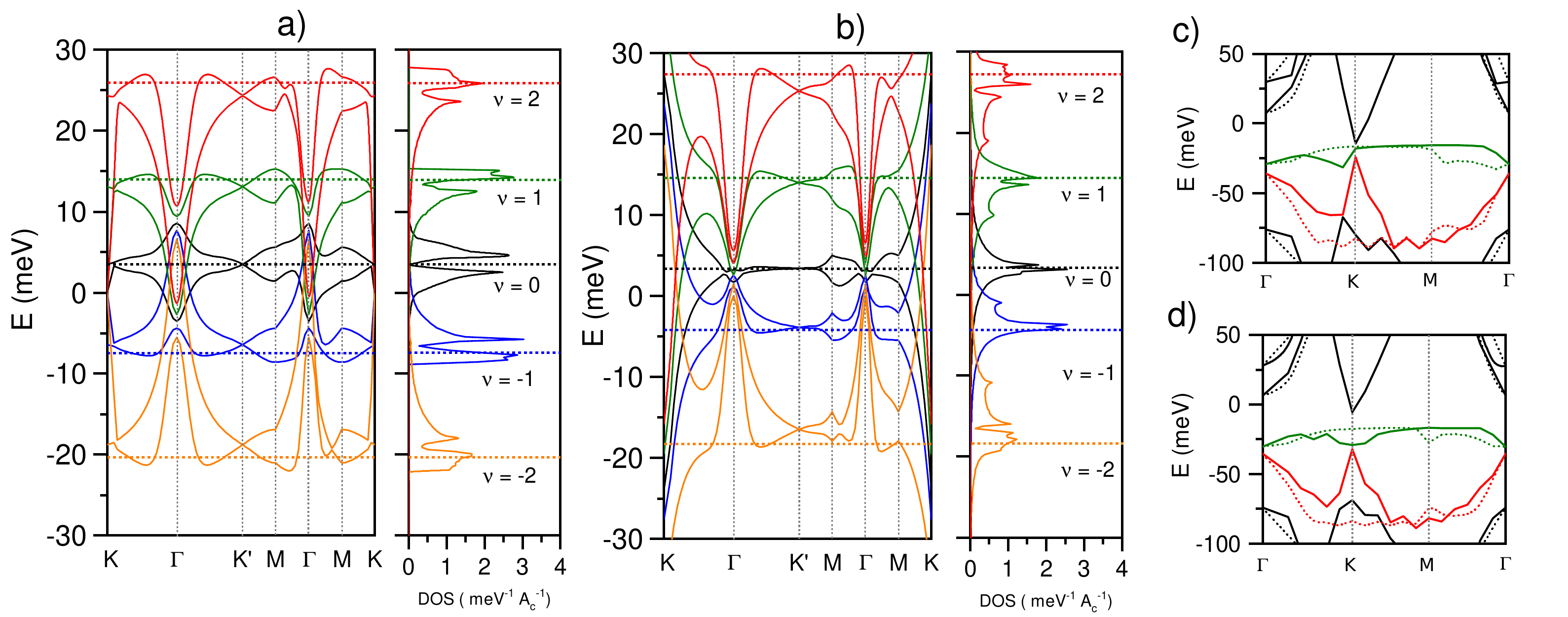}
\caption{\textbf{Hartree-Fock renormalization of band structure:} Low energy band structure and DOS obtained for $\theta=1.59^\circ$ with (a)  $\Delta V=0$ meV  and (b) $\Delta V=60$ meV. Different colors denote different fillings, as indicated in the DOS. The horizontal lines denote the Fermi level. The Fock bands at half filling are shown in for (c) $\Delta V = 0$ and (d) $\Delta V = 50$ meV. Dotted lines are the bands in the opposite valley.
}
\label{fig: figHF}
\end{figure*}

Within a single valley, we have one unbounded Dirac cone coming from the monolayer spectrum located at either $K$ or $K'$; in addition, we also have two Dirac cones at $K$ and $K'$ from the TBG's flat bands near charge neutrality. These Dirac cones are protected by $\mathcal{T} C_{2z}$ as in TBG \cite{ZP18,SW19}. The two Dirac cones from the TBG spectrum are related to each other by $\tilde{C}_{2x},$ and consequently, are pinned to the same energy. However, in the absence of particle-hole symmetry, there is no symmetry which pins the monolayer Dirac cone to the same energy. Because of this energy offset, unlike in TBG,  TTG in the single-particle limit is formally metallic for all energies in the continuum model. This important difference alters the Fermi surface geometry of TTG when compared to TBG, and may be important in the analysis of correlated insulating phases. For instance, there is technically no magic angle around which isolated energy bands become spectrally flatten like in TBG. However, the presence of $z$-mirror symmetry allows us to still define a quasi-flat band by first projecting into the mirror-odd sector and then imposing band-flatting conditions on just the two bands nearest to charge neutrality within this sector. The highly-dispersive monolayer band becomes a spectator in this scheme. For our choice of parameters, this scheme produces a magic angle at $\theta \approx 1.57^\circ$. This magic angle occurs at a larger angle than in TBG because of the enhancement in interlayer coupling strengths. In addition, we can also add an interlayer bias that couples directly to the $\ell = 1$ and $\ell = 3$ layers. The presence of this potential preserves all in-plane symmetries, but it breaks $M_z.$ Because of this, the monolayer and bilayer bands are hybridized. However, in this case, the number of Dirac cones (with the same chirality) remains preserved because the in-plane symmetries which protect them are maintained. To liberate the Dirac cones, we can add staggered sublattice potentials to the outer layers. Generically, this breaks both $M_z$ and $C_{2z}$ symmetries, and therefore, can gap out the Dirac cones as well as hybridizing the mirror-odd and mirror-even sectors. 

Band structures for  different sets of parameters are shown in Fig.~\ref{fig: Bands}. In  Fig.~\ref{fig: Bands}a, we observe that pristine TTG features decoupled bands from the mirror-odd and mirror-even sectors. A perpendicular displacement field $\Delta V = V_1-V_3$ couples these two sectors, as shown in Fig.~\ref{fig: Bands}b. The presence of a substrate induces an staggered sublattice potential that can couple the two mirror sectors as well as gapping out all the Dirac cones in the system, as shown in Fig.~\ref{fig: Bands}c. In this situation, the narrow bands are isolated from all the other bands and can carry a finite Berry curvature. Depending on the value of the staggered sublattice potential, the Chern topology of these bands can be tuned, as detailed in Ref.~\citep{SI}.   

\textit{Long-range Coulomb interaction:} At fractional fillings, interaction-driven band renormalization can significantly distort the bands represented in Fig. \ref{fig: Bands}.
We now consider the effect of the long range Coulomb interaction treated within the self-consistent Hartree-Fock approximation in Eq~(\ref{eq: Hamiltonian}). As emphasized in Refs.~\citep{Guinea2018a,Cea2019,Cea2020c} the inhomogeneous charge distribution leads to an electrostatic potential of the order of $V_{0}\sim e^{2}/\epsilon L$, where $e$ is the electron charge and $L$ the moir\'{e} length. In TBG near the magic angle, it has been shown that the electrostatic Hartree interaction strongly distorts the band structure~\citep{Cea2019}. In TTG, the situation is quite similar; we find that the self-consistent order parameter of the Hartree Hamiltonian also depends linearly on the filling fraction owing to the dominance of the narrow bands over the monolayer Dirac cones. Figure~\ref{fig: figHF} displays the band structure of TTG as a function of filling for a screening of $\epsilon = 10$. The effect of the Hartree potential varies significantly as a function of filling. This leads to an approximate pinning of the van Hove singularities at the Fermi energy at certain fillings, even in the presence of a displacement field. 

We also consider the Fock potential. In TBG, this term leads to gaps and broken symmetry phases~\cite{Cea2020c}. We analyze the system at half filling, where the Hartree potential is expected to vanish. The calculations have been carried out by projecting the Fock potential on six central bands~\cite{HF}. Results are shown in Fig.~\ref{fig: figHF}. The non-interacting flat bands are shifted downwards, and significantly widened and separated. The monolayer band is less affected, and the system remains metallic.

\textit{Superconductivity:}
Next, we study the onset of the superconducting (SC) order in TTG,
as induced by the combined effects of the 2D plasmon and
the longitudinal acoustic phonons
of the three constituting layers.
The calculation represents a straightforward generalization of the one developed by two of us
in the recent paper~\citep{cea_guinea_21} for the case of the TBG.
It consists in solving the linearized gap equation in which the pairing interaction is mediated by the Coulomb potential screened by the particle-hole excitations and the acoustic phonons within the random phase approximation (RPA).
The idea strictly recalls the Kohn-Luttinger mechanism of the superconductivity induced by repulsive interactions~\citep{KL_prl65}.

The linearized gap equation can be written as:
\begin{align}\label{Delta_tilde_eq}
\tilde{\Delta}^{m_1m_2}_{\alpha\beta}(\vec{k})=
\sum_{n_1n_2}\sum_{\vec{q}}\Gamma^{m_1m_2}_{n_1n_2;\alpha\beta}(\vec{k},\vec{q})
\tilde{\Delta}^{n_1n_2}_{\alpha\beta}(\vec{q}),
\end{align}
where $\tilde{\Delta}$ is the order parameter, $m_{1,2},n_{1,2}$ are band indices, $\alpha,\beta$ are the spin/valley flavors, and $\Gamma$ is the pairing interaction kernel,
which encodes the screened potential as well the temperature dependence of the Eq. \pref{Delta_tilde_eq}.
A detailed description of the method including the derivation of the Eq. \pref{Delta_tilde_eq} is provided in Ref. \cite{SI}.
In what follows, we consider  inter-valley superconductivity, meaning that $\alpha$ and $\beta$ have opposite valley indices in the Eq. \pref{Delta_tilde_eq}. In contrast, we do not impose any constraint to the spin texture of the SC ground state, meaning that our model can be compatible with both spin singlet or spin triplet superconductivity.
However, it is worth noting that the violation of the Pauli limit reported by the recent experiment~\citep{Cao_condmat21} strongly suggests that the SC ground state of the TTG is a spin triplet.

\begin{figure}
\includegraphics[width=3.4in]{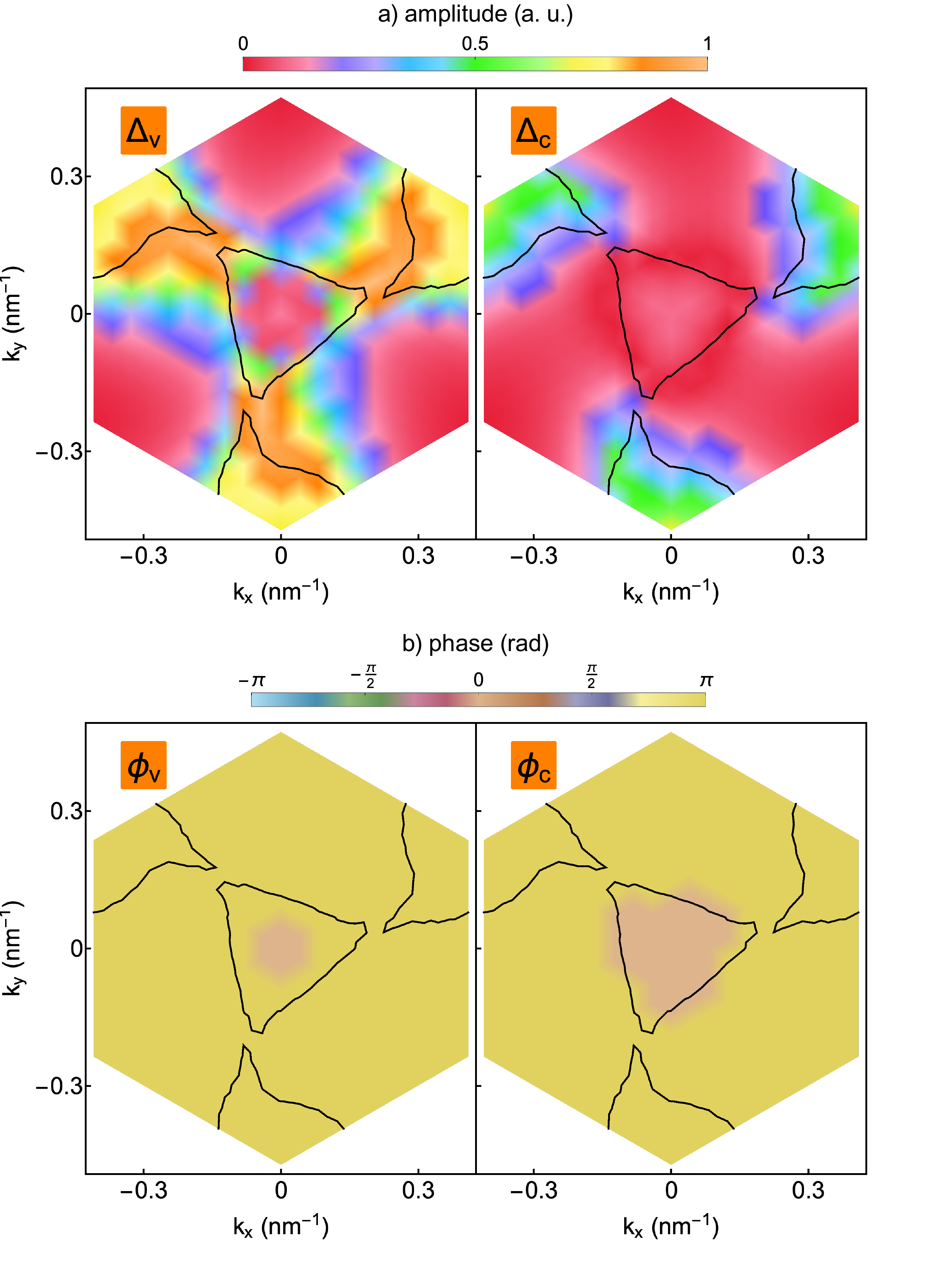}
\includegraphics[width=3.4in]{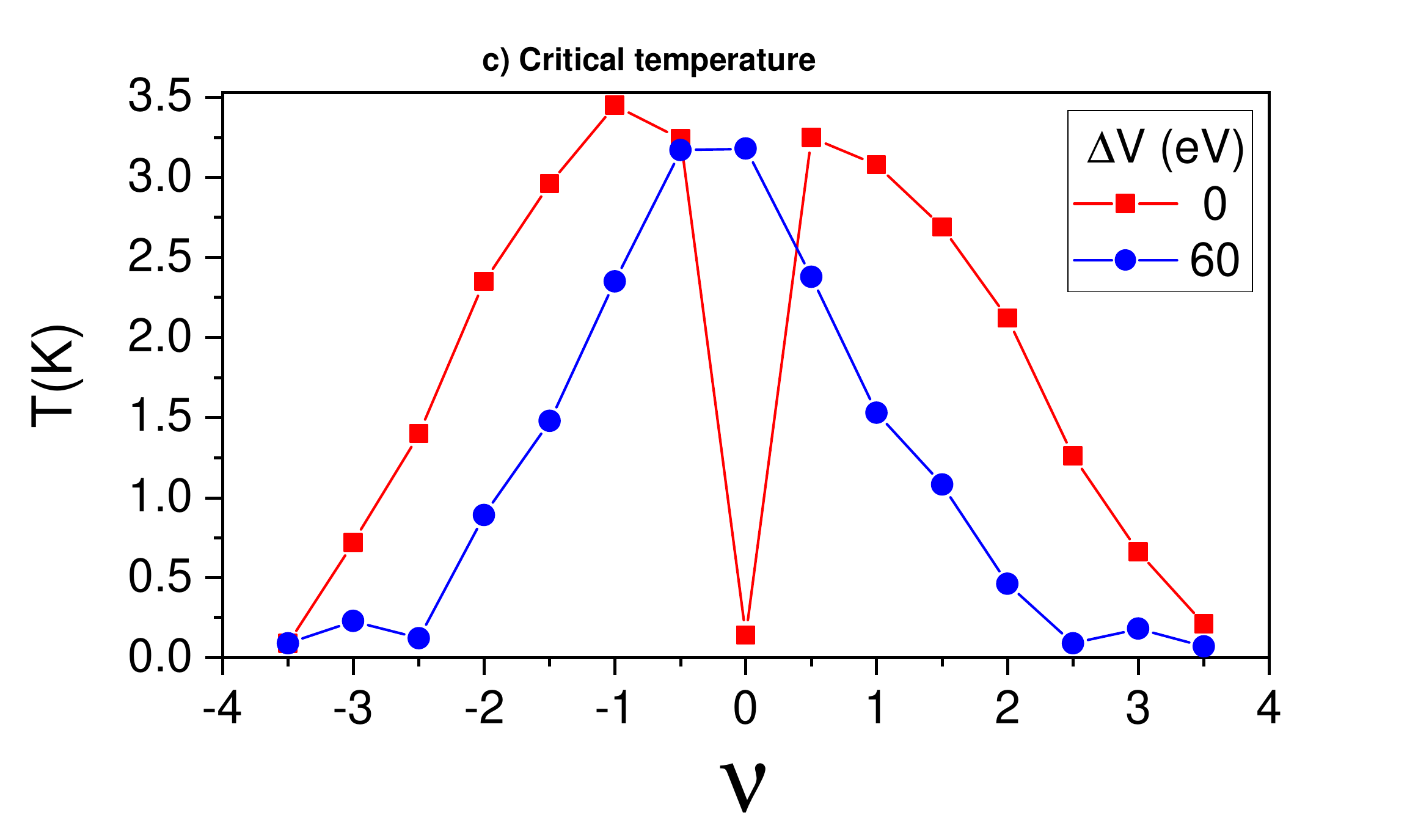}
\caption{\textbf{Superconducting order parameter and transition temperature.} Amplitude (a) and phase (b) of the SC order parameter in the valence $v$ and conduction $c$ band, as obtained for: $\theta=1.59^\circ$, $\Delta V=60$ meV and $\nu=-1$. The black lines identify the Fermi surface. c) Critical temperature as a function of the filling, $\nu$, obtained for the twist angle $\theta=1.59^\circ$ and inter-layer potential: $\Delta V=0$ meV (red line) and $\Delta V=60$ meV (blue line).}
\label{fig:OP_symmetry}
\end{figure}

Given the temperature dependence of the kernel $\Gamma$, Eq. \pref{Delta_tilde_eq} defines the SC critical temperature, $T_c$, as the one at which the largest eigenvalue of $\Gamma$ is equal to $1$.
Fig. \ref{fig:OP_symmetry} shows the critical temperature as a function of the filling, $\nu$, obtained for the twist angle $\theta=1.59^\circ$ and inter-layer potential: $\Delta V=0$eV (red line) and $\Delta V=0.06$eV (blue line). The corresponding band structure and density of states (DOS) of the central bands are shown in the Fig. \ref{fig: figHF}.
 Here, the DOS is expressed in units of meV$^{-1}A_C^{-1}$, $A_C$ being the area of the moir\'e unit cell.
 As is evident, the behavior of $T_c$ upon varying the filling follows the evolution of the DOS at the Fermi level, the latter being marked by the horizontal lines.

Finally, we study the symmetry of the SC order parameter, as given by the eigenvector of the kernel $\Gamma$, Eq.~\pref{Delta_tilde_eq}, corresponding to the eigenvalue $1$, at $T=T_c$. As a representative case,  Fig. \ref{fig:OP_symmetry} shows the amplitude (a) and the phase (b) of the order parameter, in the valence $v$ and conduction $c$ band, as obtained for: $\theta=1.59^\circ$, $\Delta V=0.06$eV and $n=-1$. The black lines identify the Fermi surface. Remarkably, also the conduction band contributes to the superconductivity, albeit marginally. The reason for that can be understood by noting that the conduction band crosses the Fermi surface close to the Dirac points, see Fig. \ref{fig: figHF}(b), where the bands flatten. The results of Fig.~\ref{fig:OP_symmetry} suggest an extended $s$-wave symmetry without sign changes along the Fermi surface, which is analogous to what obtained in the Ref.~\citep{cea_guinea_21} for the case of the TBG.

\textit{Conclusions:}
We have analyzed the symmetries, band structure, role of interactions, and superconductivity of twisted trilayer graphene near the first magic angle. Our results are similar to those found for twisted bilayer graphene. The presence of an additional layer, and the accompanying electronic bands with monolayer dispersion, does not modify qualitatively the features found in twisted bilayer graphene.

To summarize, our main results are:

- The long-range electron-electron interaction modifies significantly the central bands. The Hartree potential widens the bands, leading to a bandwidth $\sim 20-40$ meV. The exchange potential at half filling separates the occupied valence band from the empty conduction band, which become $20 -40$ meV apart. The presence of an additional linearly dispersive band prevents the formation of a gap and an insulating state.

- Long-wavelength charge fluctuations, renormalized by their coupling  by the electrostatic interaction, and also by their interaction with longitudinal acoustic phonons, are sufficient to induce superconductivity, with critical temperatures of up to a few Kelvin. The order parameter has a significant structure near the Fermi surface pockets. Pairing by long-wavelength excitations, which do not mix valleys, leads to degenerate spin singlet/valley triplet and spin triplet/valley singlet  superconducting phases.

\textit{Acknowledgements.}
This work was supported by funding from the European Commision, under the Graphene Flagship, Core 3, grant number 881603, and by the grants NMAT2D (Comunidad de Madrid, Spain),  SprQuMat and SEV-2016-0686, (Ministerio de Ciencia e Innovación, Spain). VTP acknowledges support from the NSF Graduate Research Fellowships Program and the P.D. Soros Fellowship for New Americans.

\bibliography{main}

\begin{thebibliography}{51}
\bibitem [{HF()}]{HF}%
  \BibitemOpen
  \href@noop {} {}\bibinfo {note} {We assume that the initial hoppings include
  the renormalization due to the exchange potential from the bands other than
  the sisx central bands, so that no substraction of the exchange potential is
  required.}\BibitemShut {Stop}%
\end{thebibliography}

\appendix
\onecolumngrid
\setcounter{equation}{0}
\setcounter{figure}{0}
\renewcommand{\theequation}{S\arabic{equation}}
\renewcommand{\thefigure}{S\arabic{figure}}
\renewcommand{\bibnumfmt}[1]{[#1]}
\renewcommand{\citenumfont}[1]{#1}

\pagebreak

\begin{center}
\begin{Large}
Supplementary Material
\end{Large}
\end{center}

\subsection{Construction of Hamiltonian}
Here, we study twisted trilayer graphene in which all three layers are rotated with respect to adjacent layers. Before proceeding to the analysis of multilayer systems, we first establish the convention of monolayer graphene.  We choose to align the $p_z$ orbitals on both sublattices so that they have the same sign convention. In this case, $-\gamma_0,$ the hopping integral between in-plane nearest neighbors, is a negative constant (we have chosen a minus sign so that $\gamma_0$ is positive). Let $a$ denote the lattice constant,  $\mathbf{a}_1 = a\left( 1, 0\right),$ and $\mathbf{a}_2 = a\left( \frac{1}{2}, \frac{\sqrt{3}}{2}\right).$ The primitive reciprocal lattice vectors $\mathbf{g}_1 = \frac{2 \pi}{a} \left( 1,-\frac{1}{\sqrt{3}} \right)$ and $\mathbf{g}_2 = \frac{2 \pi}{a} \left(  0, \frac{2}{\sqrt{3}}  \right).$ The zone corners are located at $\mathbf{K}_\pm = \frac{4 \pi}{3  a} \left(  \mp 1,0 \right).$ The minimal tight-binding model in the basis of  $\ket{A}$ and $\ket{B}$ is 
\begin{equation}
\mathfrak{h}(\mathbf{k}) = \begin{pmatrix}
0 & -\gamma_0\phi(\mathbf{k}) \\
-\gamma_0 \phi(\mathbf{k})^*  & 0
 \end{pmatrix},
\end{equation}
where $\phi(\mathbf{k}) =  \sum_{j = 1}^3 e^{i \mathbf{k} \cdot \boldsymbol{\delta}_j}$ and $\boldsymbol{\delta}_j = \frac{a}{\sqrt{3}} \left( \sin \left( \frac{2 \pi j}{3 }\right), -\cos \left( \frac{2 \pi j}{3 } \right) \right).$ Expanding about the zone corners, we find
\begin{equation}
\begin{split}
\mathfrak{h}\left(\mathbf{p} = \mathbf{k}- \mathbf{K}_- \right) &= \begin{pmatrix}
0 & \frac{\sqrt{3}  }{2}a\gamma_0 \left( p_x+ip_y \right)\\
\frac{\sqrt{3}  }{2}a\gamma_0 \left( p_x-ip_y \right) & 0
\end{pmatrix} = -\frac{\sqrt{3}}{2} a\gamma_0 \mathbf{p} \cdot \left( - \sigma_x, \sigma_y \right) ,\\
\mathfrak{h}\left( \mathbf{p} = \mathbf{k}- \mathbf{K}_+ \right) &=\begin{pmatrix}
0 & -\frac{\sqrt{3}  }{2}a\gamma_0 \left( p_x-ip_y \right)\\
-\frac{\sqrt{3}  }{2}a\gamma_0 \left( p_x+ip_y \right) & 0
\end{pmatrix} = -\frac{\sqrt{3}}{2}a \gamma_0 \mathbf{p} \cdot \left( + \sigma_x, \sigma_y \right).
\end{split}
\end{equation}
Defining $\hbar v_F = \sqrt{3}a\gamma_0/2,$ we can summarize as follows: $\mathfrak{h} \left( \mathbf{k} \right) = -\hbar v_F \left( \mathbf{k}- \mathbf{K}_\nu \right) \cdot \left( \nu \sigma_x, \sigma_y \right).$

Next, we consider stacking up three layers with a relative twist between them. We denote a rotation matrix of the form 
\begin{equation}
R(\theta) = \begin{pmatrix}
\cos \left(\theta \right)  & -\sin \left( \theta \right) \\
\sin \left( \theta \right) & \cos \left( \theta \right)
\end{pmatrix}.
\end{equation}
This performs a counter-clockwise rotation. We consider three layers originally-aligned in $AA$ configuration. We then rotate layer $\ell = 1$ by $-\theta/2,$ layer $\ell = 2$ by $\theta/2,$ and layer $3$ by $-\theta/2.$ The primitive lattice vectors are rotated by $\mathbf{a}^\ell_i = R(\pm \theta/2) \mathbf{a}_i$ and the primitive reciprocal lattice vectors by $\mathbf{g}^\ell_i = R(\pm \theta/2) \mathbf{g}_i.$ The primitive moir\'{e} reciprocal lattice vectors are then given by 
\begin{equation}
\mathbf{G}_1^\text{M} = \mathbf{g}^1_1-\mathbf{g}^2_1 = G_\text{M} \left( - \frac{1}{2},-\frac{\sqrt{3}}{2} \right), \quad \mathbf{G}_2^\text{M} = \mathbf{g}^1_2-\mathbf{g}^2_2 =  G_\text{M} \left( - 1,0  \right).
\end{equation}
where $G_\text{M}  = 8 \pi \sin \left( \theta/2\right)/(\sqrt{3}a).$ The primitive moir\'{e} lattice vectors
\begin{equation}
\mathbf{L}_1^\text{M} =L_\text{M} \left( 0, -1 \right), \quad \mathbf{L}_2 = L_\text{M} \left( - \frac{\sqrt{3}}{2},  \frac{1}{2}\right),
\end{equation}
$L_\text{M} = a/(2 \sin \left( \theta/2 \right)).$ The basic valley-projected Hamiltonian in mixed real-space and reciprocal-space representation is 
\begin{equation}
\label{eq: 6}
\mathcal{H}^{\theta}_{\nu} (\mathbf{k}) = \begin{pmatrix}
\mathfrak{h} \left( \mathbf{k}^{+\frac{\theta}{2}} \right) &  \left[\mathfrak{t}_\nu^\theta (\mathbf{r}) \right]^\dagger & 0\\
\mathfrak{t}_\nu^\theta (\mathbf{r}) &\mathfrak{h} \left( \mathbf{k}^{-\frac{\theta}{2}}  \right) &  \mathfrak{t}_\nu^\theta (\mathbf{r}) \\
0 & \left[ \mathfrak{t}_\nu^\theta (\mathbf{r}) \right]^\dagger & \mathfrak{h} \left( \mathbf{k}^{+\frac{\theta}{2}}  \right) 
\end{pmatrix},
\end{equation}
where $\mathbf{k}^{(\pm\theta/2)} = R\left( \pm \theta/2 \right) \mathbf{k}$ (note the convention that layer 1 is rotated by $-\theta/2$  so the wavevector must be rotated by $+\theta/2$ in the opposite direction to compensate), and 
\begin{equation}
\mathfrak{t}_\nu^\theta (\mathbf{r})  = \begin{pmatrix}
w_0 & w_1 \\
w_1 & w_0
\end{pmatrix} + \begin{pmatrix}
w_0 & w_1 e^{-2 \pi \nu i /3} \\
w_1 e^{2 \pi \nu i /3}& w_0
\end{pmatrix} e^{i \nu \mathbf{G}_1^\text{M} \cdot \mathbf{r}}  + \begin{pmatrix}
w_0 & w_1 e^{2 \pi \nu i /3} \\
w_1 e^{-2 \pi \nu i /3}& w_0
\end{pmatrix} e^{i \nu \left(\mathbf{G}_1^\text{M}+\mathbf{G}_2^\text{M} \right) \cdot \mathbf{r}}.
\end{equation}
The Hamiltonian in Eq. \eqref{eq: 6} is in the ``natural" basis of sublattices and layers. As shown in Ref. \cite{KKTV19}, we can perform a unitary transformation to simplify the Hamiltonian
\begin{equation}
\begin{split}
\tilde{\mathcal{H}}_\nu^\theta(\mathbf{k}, \mathbf{r}) = &\begin{pmatrix}
\frac{1}{\sqrt{2}} & 0 & \frac{1}{\sqrt{2}} \\
0 & 1 & 0 \\
-\frac{1}{\sqrt{2}} & 0 & \frac{1}{\sqrt{2}} 
\end{pmatrix}\begin{pmatrix}
\mathfrak{h} \left( \mathbf{k}^{+\frac{\theta}{2}}  \right) &  \left[\mathfrak{t}_\nu^\theta (\mathbf{r}) \right]^\dagger & 0\\
\mathfrak{t}_\nu^\theta (\mathbf{r}) &\mathfrak{h} \left( \mathbf{k}^{-\frac{\theta}{2}}  \right) &  \mathfrak{t}_\nu^\theta (\mathbf{r}) \\
0 & \left[ \mathfrak{t}_\nu^\theta (\mathbf{r}) \right]^\dagger & \mathfrak{h} \left( \mathbf{k}^{+\frac{\theta}{2}}  \right) 
\end{pmatrix} \begin{pmatrix}\frac{1}{\sqrt{2}} & 0 & -\frac{1}{\sqrt{2}} \\
0 & 1 & 0 \\
\frac{1}{\sqrt{2}} & 0 & \frac{1}{\sqrt{2}} 
\end{pmatrix}\\
&= \begin{pmatrix}
\mathfrak{h} \left( \mathbf{k}^{+\frac{\theta}{2}}  \right) &  \sqrt{2}\left[\mathfrak{t}_\nu^\theta (\mathbf{r}) \right]^\dagger & 0\\
\sqrt{2}\mathfrak{t}_\nu^\theta (\mathbf{r}) &\mathfrak{h} \left( \mathbf{k}^{-\frac{\theta}{2}} \right) &  0 \\
0 & 0 & \mathfrak{h} \left( \mathbf{k}^{+\frac{\theta}{2}}\right) 
\end{pmatrix}.
\end{split}
\end{equation}
The new basis states mix layers as follows
\begin{equation}
\begin{pmatrix}
\ket{\mathcal{M}_\text{o}^1} \\
\ket{\mathcal{M}_\text{o}^2} \\
\ket{\mathcal{M}_\text{e}}
\end{pmatrix} = \begin{pmatrix}\frac{1}{\sqrt{2}} & 0 & \frac{1}{\sqrt{2}} \\
0 & 1 & 0 \\
-\frac{1}{\sqrt{2}} & 0 & \frac{1}{\sqrt{2}} 
\end{pmatrix}\begin{pmatrix}
\ket{\ell = 1} \\
\ket{\ell =2} \\
\ket{\ell =3}
\end{pmatrix} = \begin{pmatrix}
\frac{1}{\sqrt{2}} \ket{\ell =1} + \frac{1}{\sqrt{2}} \ket{\ell =3}  \\
\ket{\ell =2} \\
\frac{1}{\sqrt{2}} \ket{\ell =3} - \frac{1}{\sqrt{2}} \ket{\ell =1}
\end{pmatrix}.
\end{equation}

\begin{figure}
\includegraphics[width=4in]{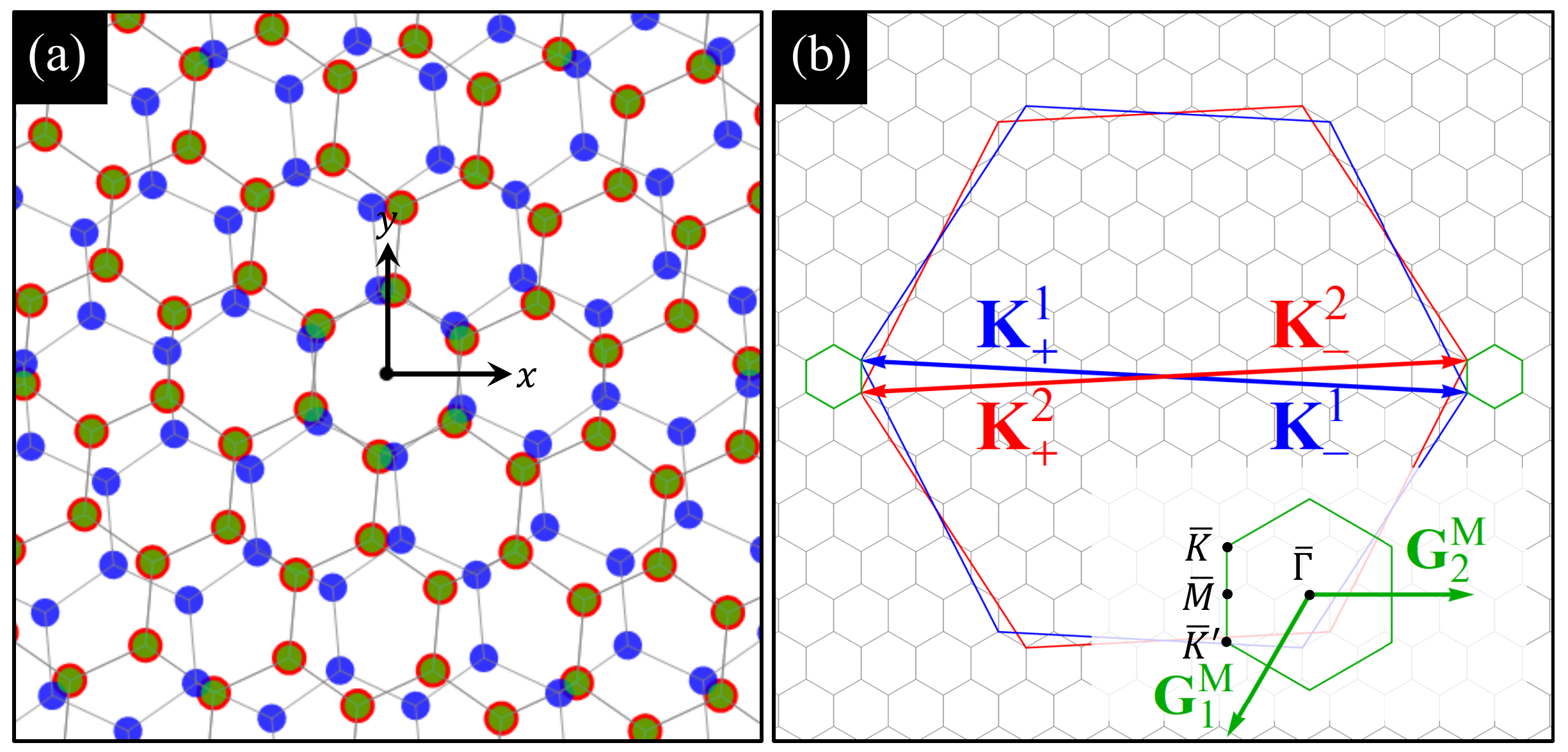}
\caption{\textbf{Representation of twisted trilayer graphene in real and reciprocal space.} (a) Carbon atoms from layers 1, 2, and 3 are shaded red, blue, and green respectively. We note that the red and green circles often coincide because if the middle layer were not present, the top and bottom layers would be right on top of one another. (b) Moiré Brillouin zone is shown in green with high-symmetry points indicated. The microsopic graphene Brillouin zones are shown in blue, for layer 1, and red, for layer 2. For clarity, we do not show explicitly the Brillouin zone for layer 3 since it is identical to that of layer 1. The high-symmetry points in the moiré Brillouin zone are denoted with bars for emphasis here to distinguish them from the microscopic high-symmetry points; these bars are dropped in the main text for simplicity. }
\label{fig: structure}
\end{figure}


\subsection{Symmetry-Breaking Perturbations} 

The appeal of TTG is its robust tunability. As such, we now consider some ways through which such tunability can be controlled. The most generic spatially-uniform perturbation we consider takes the form, in the two different bases,
\begin{equation}
\label{eq: perturbation}
\mathcal{V} = \begin{pmatrix}
V_1 & 0 & 0 \\
0 & 0 & 0 \\
0 & 0 & V_3
\end{pmatrix} \quad \text{and} \quad \tilde{\mathcal{V}} = \frac{1}{2} \begin{pmatrix}
V_1 + V_3 & 0 & V_3-V_1 \\
0 & 0 & 0 \\
V_3-V_1 & 0 & V_1+V_3
\end{pmatrix},
\end{equation}
where $V_1$ and $V_3$ are $2 \times 2$ matrices that act on the sublattice degrees of freedom. We do not consider perturbations that directly couple to  the middle layer since it is encapsulated and is presumably difficult to practically access. For $V_1 = V_3,$  $z$-mirror symmetry is preserved; otherwise, the mirror-odd and mirror-even sectors are mixed by $V_3-V_1.$

There are two limits to consider: $V_1 = + V_3$ and $V_1=-V_3$. The former retains the decoupled bilayer-monolayer form because the off-diagonal terms vanish.  In this case, we can still consider the composite system as two separate systems: one monolayer and one twisted bilayer graphene. However, both the bilayer system and the monolayer system may have additional perturbations that may lower their symmetries. In the latter situation, the bilayer and monolayer systems are mixed via the off-diagonal terms. An example of the second limit is the  displacement field, $\pm \Delta V/2$, between the outermost layers.  The generic system varies between this two limits. In addition to the extrinsic perturbations discussed above, there is also an intrinsic perturbation to Eq. \eqref{eq: 6} in the original layer basis of the form
\begin{equation}
\begin{split}
\mathcal{V}_{AA} = \begin{pmatrix}
0 & 0 & \mathfrak{t}_\text{AA} \\
0 & 0 & 0 \\
\mathfrak{t}_\text{AA}^\dagger  & 0 & 0
\end{pmatrix},  \quad 
\mathfrak{t}_\text{AA} = \begin{pmatrix}
\gamma_2 & 0 \\
0 & \gamma_2
\end{pmatrix},
\end{split}
\end{equation}
where $\gamma_2$ is the $AA$ coupling constant of the first and third layers. Because these two layers are  far from each other, $\gamma_2$ is presumably a very small parameter compared to the other energy scales in the problem. To estimate, we use the conventional Slater-Koster parameterization \cite{KYK18}
\begin{equation}
\gamma_2(r) = V_{pp\sigma}^0 \exp \left( \frac{d_0-r}{r_0} \right).
\end{equation}
We take $V_{pp\sigma}^0 = 0.48$ eV, $d_0 = 0.335$ nm, and $r_0 = 0.045$ nm. We take $r = 2d_0$ to find $\gamma_2 \approx 0.3$ meV, indeed a very small value as expected. However, as noted in the main text, this may be an underestimate of $\gamma_2$ since it assumes that this parameter is environment-independent.

\begin{figure*}
\includegraphics[scale=0.60]{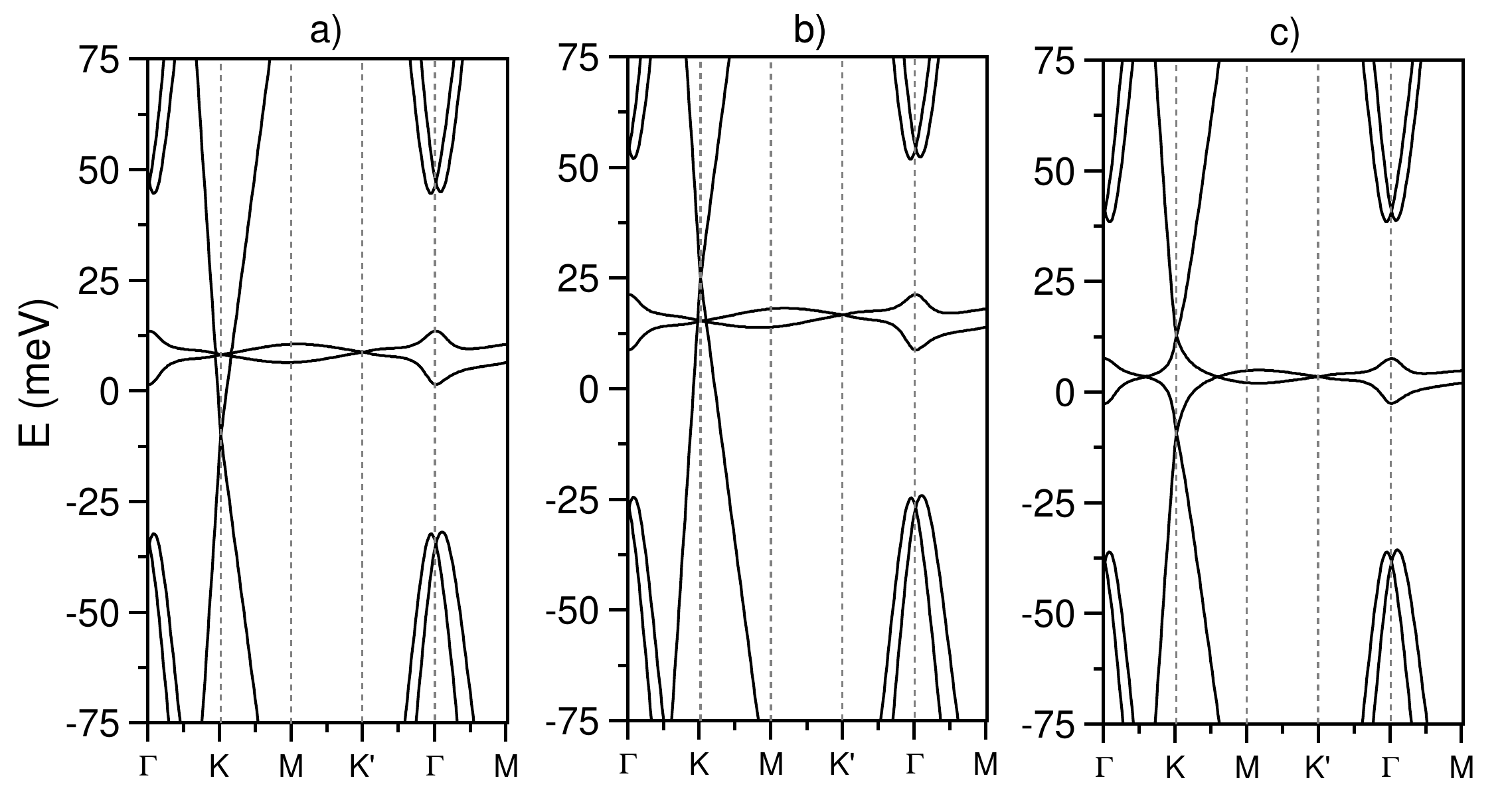}
\caption{\textbf{Symmetry-Breaking Perturbations:} a) Plot of TTG with a twist angle of $\theta = 1.59^\circ$ and $\gamma_2 = 10 $ meV. b) For $V_1=V_3 = 2\Delta \tilde{V} = 10 $ meV the Dirac cone is shifted but is still decoupled from the narrow bands. c) For $V_1=-V_3$, both Dirac cones are coupled to the narrow bands. This situation is equivalent to a displacement field, $\Delta V$, between the outermost layers. In the figure we use $2\Delta V = 10 $ meV.}
\label{fig: perturbations}
\end{figure*}

\subsection{Band Structure}

We now study the band structure of TTG.  In the intrinsic case, $z$-mirror symmetry allows us to partition the Hamiltonian into a direct sum of two Hamiltonians: one describing a twisted graphene bilayer  with enhanced interlayer coupling by a factor of $\sqrt{2}$ and one describing a simple  graphene monolayer. Therefore, much of the band structure analysis of this case is the same as the analysis of TBG. The monolayer Dirac cone is simply a spectator here. The system as a whole is metallic for all energies in this continuum construction without particle-hole symmetry. However, once we project onto just the mirror-odd sector, we can still speak of flat bands within just this sector of the Hamiltonian. This is useful for later because once we turn on symmetry-breaking perturbations that hybridize the mirror-odd and mirror-even sectors, some degeneracies will be gapped out, and true isolated flat bands can emerge. Importantly, the flatness of these bands is due primarily to the flatness of the parent bands that descend from the mirror-odd sector. Thus, focusing for a moment on just the mirror-odd sector, $\ket{\mathcal{M}_\text{o}^1}$ and $\ket{\mathcal{M}_\text{o}^2},$ we first establish regions of parameter space where flat bands emerge.

\begin{figure}
\includegraphics[scale=0.40]{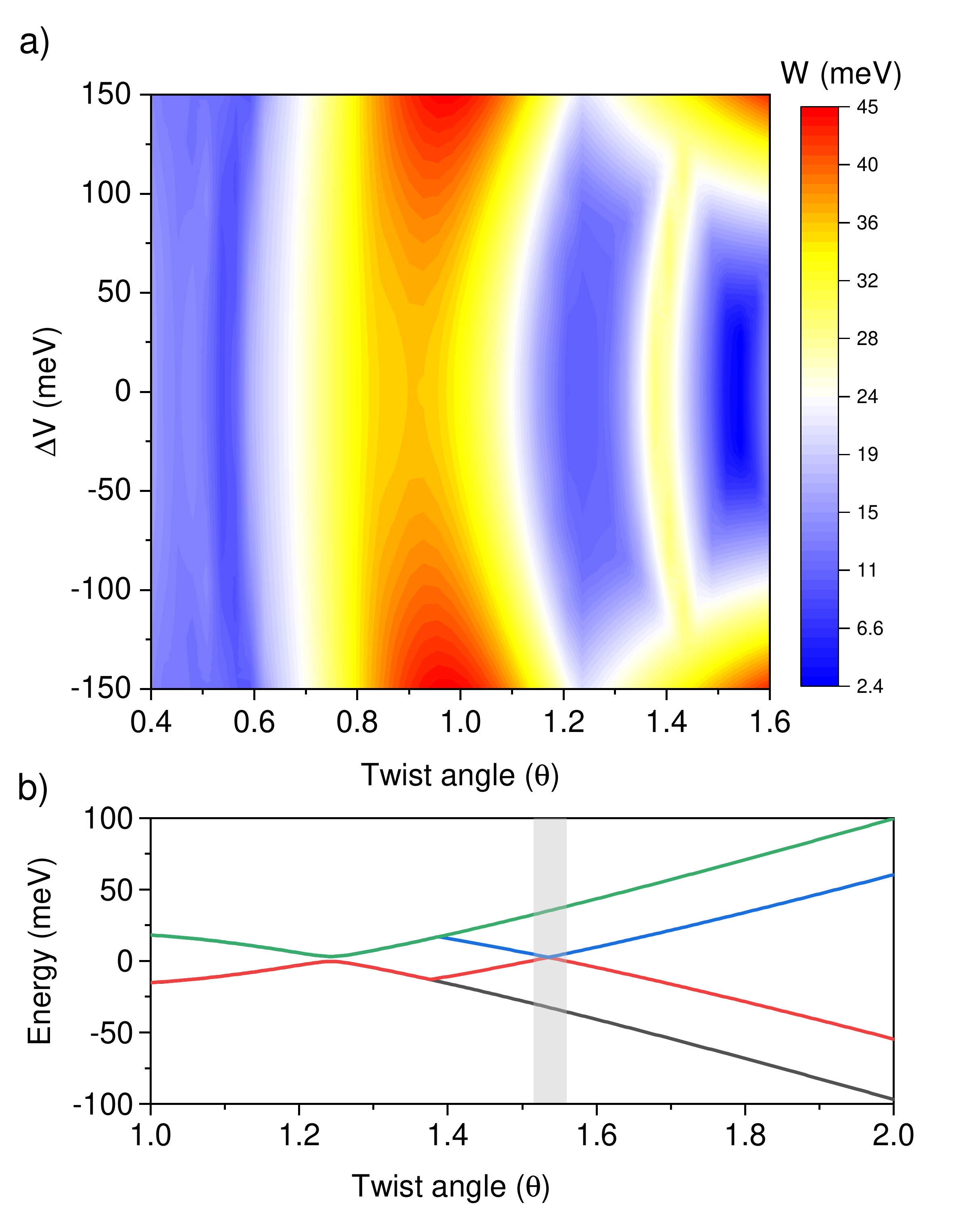}
\caption{\textbf{Magic-angle region:} The top plot shows the bandwidth, $W$, as defined in the text, as a function of angle and displacement field. Bottom plot shows the four energy eigenvalues closest to charge neutrality of the mirror-odd bands at $\bar{\Gamma}.$ The band gaps between the middle two flat bands and the higher-energy bands close when the top two or bottom two eigenvalues coincide. Region in $\theta$ where magic angle occur is shaded in light gray.}
\label{Fig magic}
\end{figure}

By a magic angle, we mean an angle at which a pair of entangled flat bands emerges near charge neutrality; these flat bands are isolated from the other energy bands. This definition does not produce unique magic angles because the criteria for flatness remain ambiguous. One common definition is to pin magic angles at those at which the Dirac velocity at $\bar{K}$ and $\bar{K}'$ vanishes in all directions. However, this definition does not always coincide with small bandwidth. Indeed, vanishing of the Dirac velocity does not even imply band isolation from the other high-energy bands. For us, we instead use the local minimalization of the bandwidth as the definition of magic angle. To find the magic angle, we find regions of $\theta$ where isolated bands can occur. This can be  found numerically by computing the four lowest energy eigenvalues near charge neutrality at $\bar{\Gamma}.$ When the top two or bottom two eigenvalues coincide, this means that a gap cannot exist. Then, for regions where a gap could exist, we compute the corresponding bandwidths, defined as the maxima of the energies computed at high-symmetry points $\bar{\Gamma},$ $\bar{K},$ $\bar{M},$ and $\bar{M}'$ minus the minima. These may, of course, not be the true bandwidths, but they are sufficient to identify the magic angle. Using this condition we find a magic angle around $\theta \approx 1.57^\circ$, as shown in Fig.~\ref{Fig magic}. We note that due to the enhancement of the interlayer coupling, the magic angle, within the window of parameters in Fig.~\ref{Fig magic}b, occurs at a larger value than that known for TBG. In addition, this magic angle is persistent even in the presence of a displacement field, as shown in Fig.~\ref{Fig magic}a. 


\subsection{Sublattice potential and topological phases}

\begin{figure}
\includegraphics[scale=0.30]{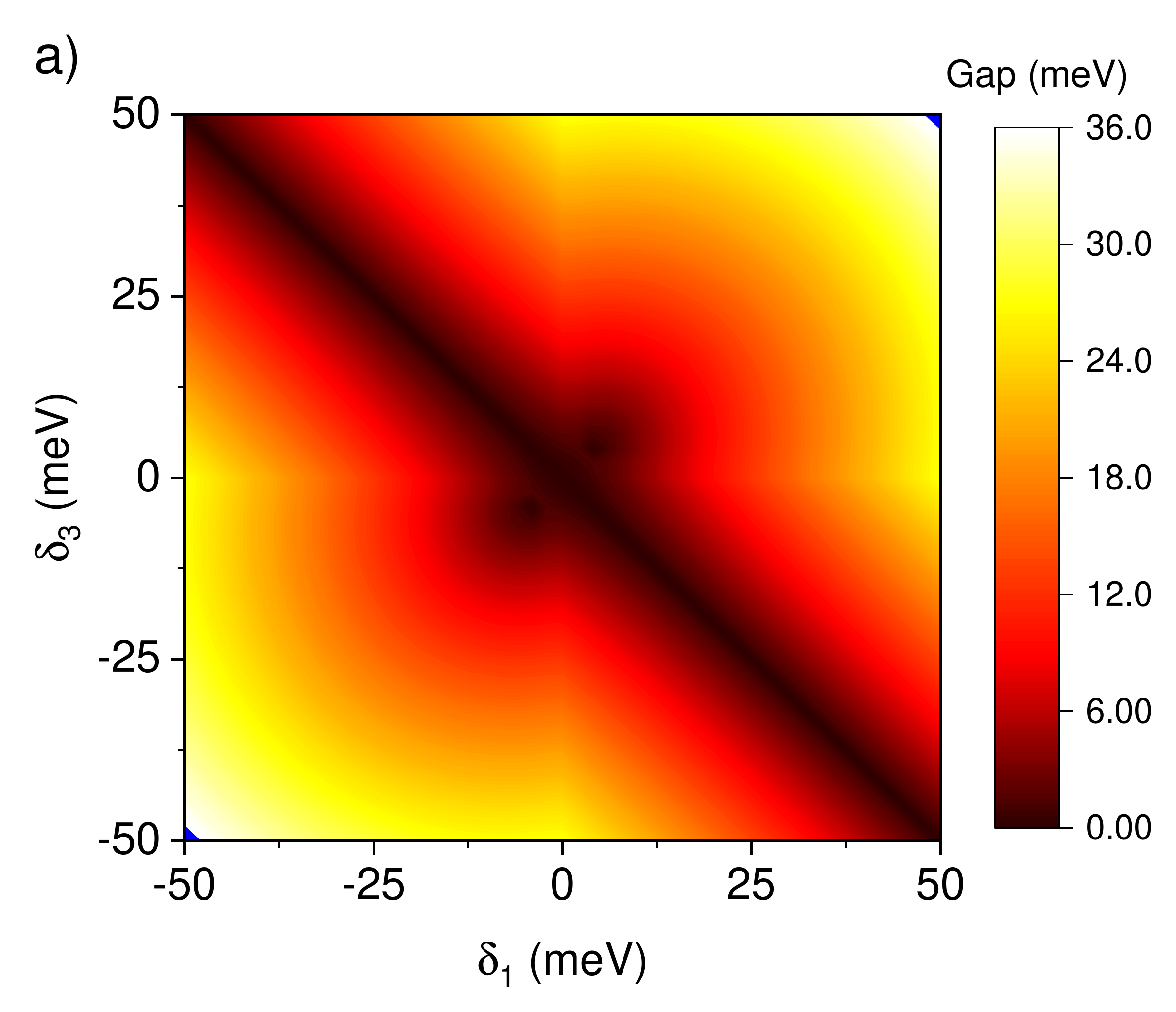} \includegraphics[scale=0.30]{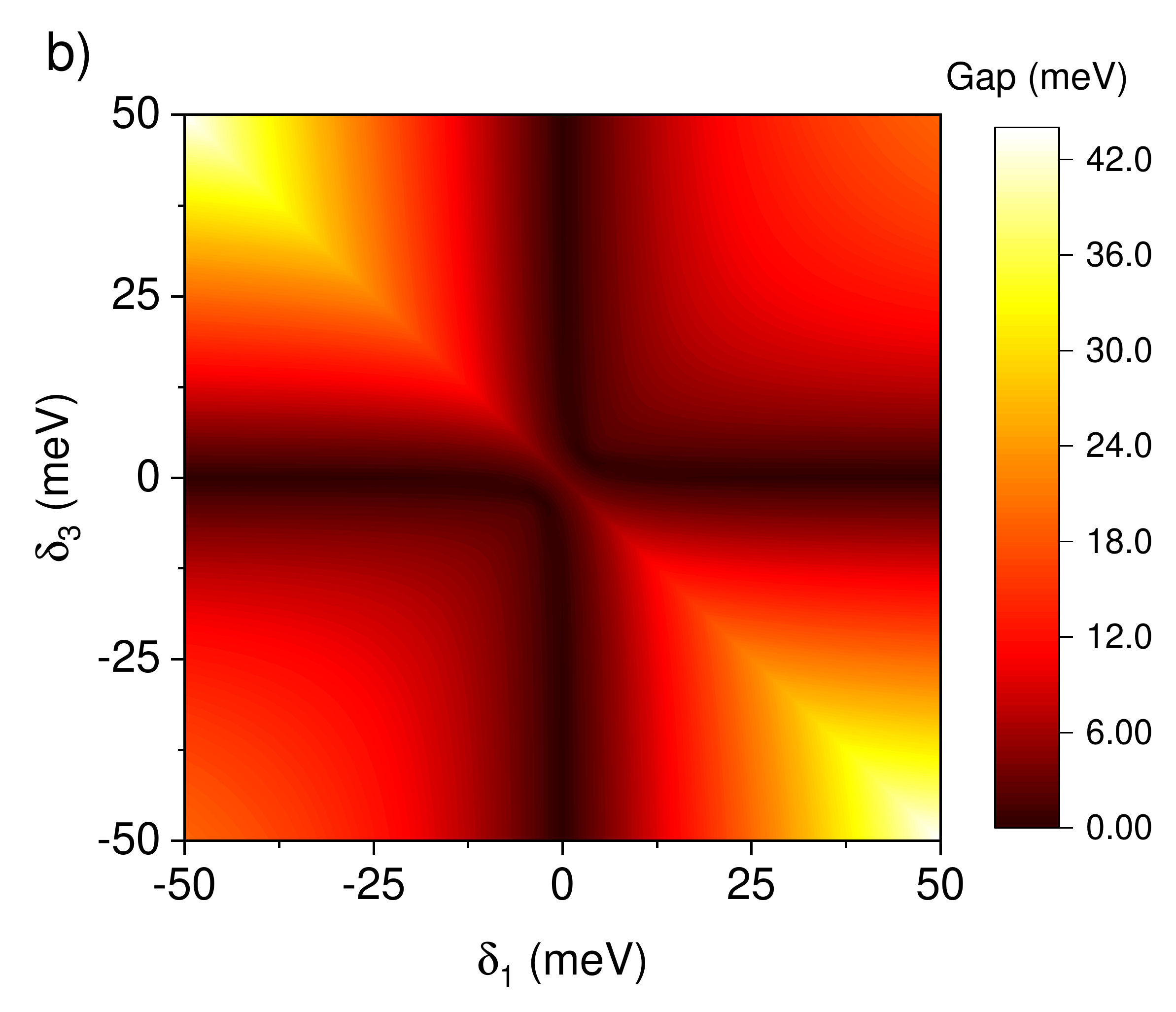} \includegraphics[scale=0.30]{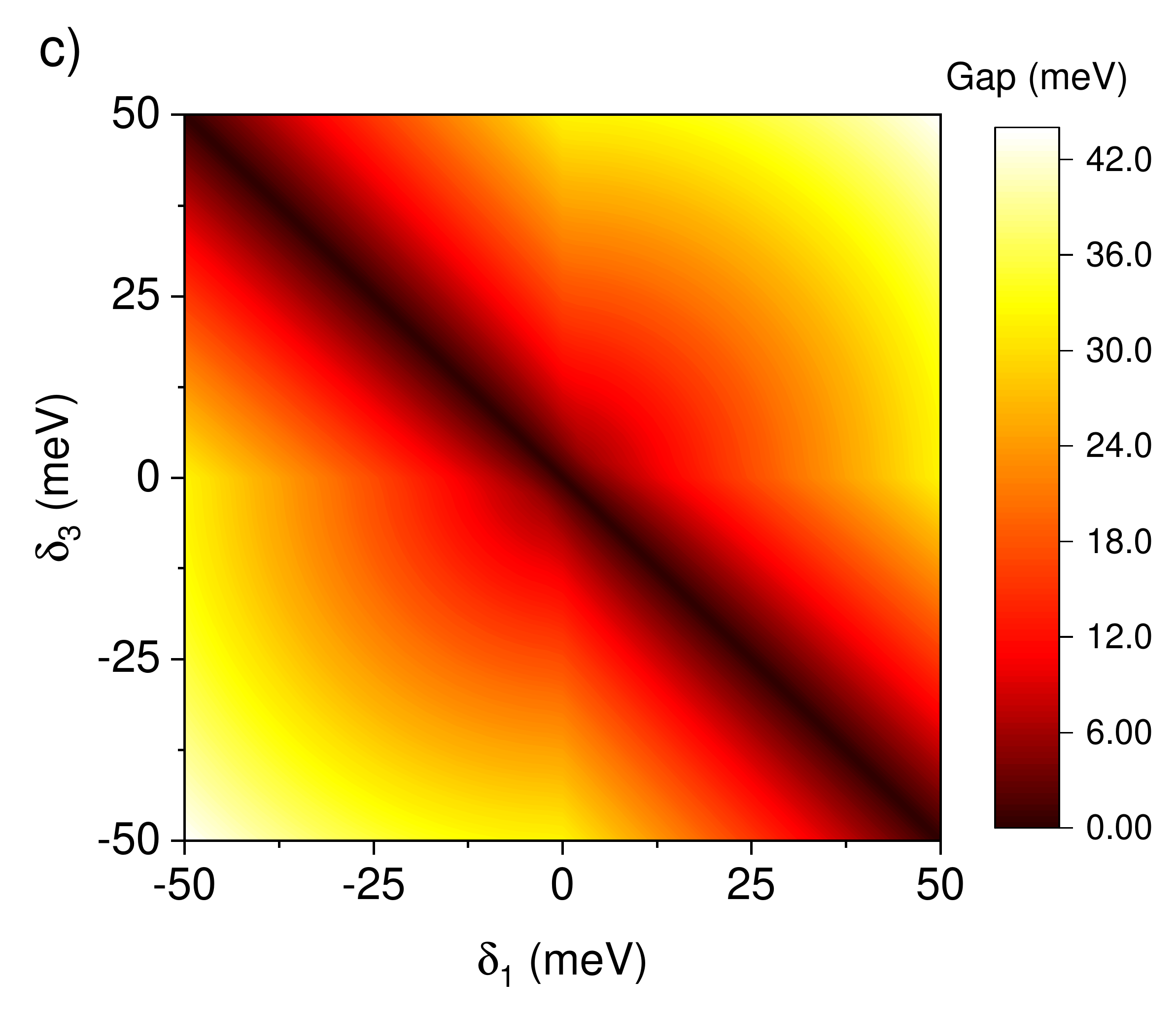}   \includegraphics[scale=0.30]{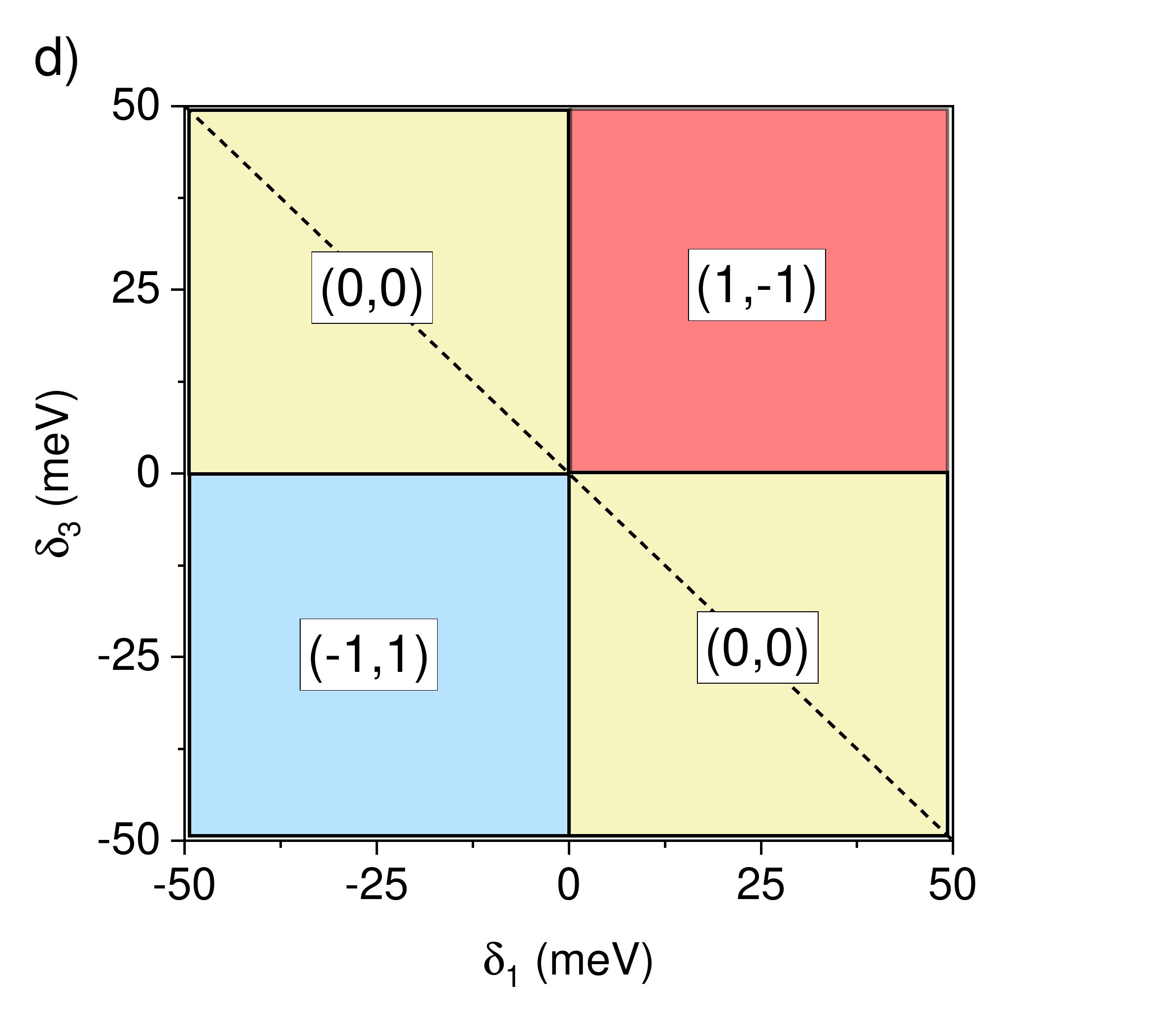}
\caption{\textbf{Mass gap:} At the point $\vec K$ and as a function of the mass terms we calculate the gap between a) the upper Dirac cone and the top narrow band, b) top and bottom narrow bands and c) bottom narrow band and lower Dirac cone. In d) we show the topological phases as a function of the staggered sublattice potential. Continuous lines are the boundaries in which there is a topological phase transition. Dashed diagonal line separates the region within the same topological phase where the signs of the sublattice potential are inverted. 
}
\label{Fig gaps}
\end{figure}

We now consider the presence of an staggered sublattice potential. As in the case of TBG, a mass gap can be induced by the presence of a substrate~\citep{Cea2020TBGhBN,Shi2021tBGhBN,Shin2021tBGhBN,Mao2021tBGhBN,Lin2021tBGhBN}. For TTG, we consider a mass gap of the form $\delta_{l}\sigma_{z}$ with $\ell=\{1,3\}$ labelling the corresponding term acting on the bottom and top layer. As shown in Fig.~\ref{fig: Bands}c, a staggered sublattice potential $\delta_{1} = \delta_{3} = 10$ meV opens a gap between the narrow bands and the Dirac cones. Fig.~\ref{Fig gaps} shows a density plot of the magnitude of the gap between the narrow bands and the Dirac cones as a function of the mass terms $\delta_{1/3}$. Fig.~\ref{Fig gaps}a is the gap between the upper Dirac cone and the top narrow band, Fig.~\ref{Fig gaps}b is the gap between the narrow bands and Fig.~\ref{Fig gaps}c the gap between the lower narrow band and the lower Dirac cone.  Some effects are in order here; in Fig.~\ref{Fig gaps}a and Fig.~\ref{Fig gaps}c, we notice that if $\delta_{1}=-\delta_{3}$, each narrow band is connected to a Dirac cone at $K$ and connected to each other at $K'$. In other regions of the same figure, the narrow bands are decoupled from the cones with a finite gap. However, as shown in the dark regions in Fig.~\ref{Fig gaps}b if any $\delta_{1}$ or $\delta_{3}$ is zero, the narrow bands are decoupled from the Dirac cones but with no gap between them. In Fig.~\ref{Fig gaps}d we display the topological phases as a function of the staggered sublattice potential. In the region where both $\delta_{1}$ and $\delta_{3}$ are negatives both narrow bands are isolated with Chern numbers $\left( \mathcal{C}_{1},\mathcal{C}_{2} \right)=\left(-1,1\right)$, if both are positives the phase is $\left(1,-1\right)$. The trivial phase $\left(0,0\right)$ is when both mass terms have different sign. 

\subsection{Self-consistent Hartree Interaction}

In  momentum space, the self-consistent Hartree interaction has matrix elements given by \citep{Guinea2018a,Cea2019}
\begin{equation}
\rho_{H}(\vec G)=4V_{C}\left(\vec G\right) \int_{BZ}\frac{d^{2}\boldsymbol{k}}{V_{mBZ}}\sum_{\vec G^{\prime},l}\phi_{k,l}^{\dag}\left(\vec G^{\prime}\right)\phi_{k,l}\left(\vec G+\vec G^{\prime}\right),\label{eq: rho components} 
\end{equation}
where $V_{C}\left(\vec G\right)=\frac{2\pi e^{2}}{\epsilon \left|\vec G\right|}$ is the Fourier transform of the Coulomb potential evaluated at $\vec G$, $V_{mBZ}$ is the area of the moir\'{e} Brillouin zone, and the factor 4 takes into account spin/valley degeneracy. The parameter $l$ is a band index resulting from the diagonalization of the full Hamiltonian in Eq.~(\ref{eq: Hamiltonian}). In the above equation, $\phi_{k,l}\left(\vec G \right)$ is the amplitude for an electron to occupy a state with momentum $\vec k+\vec G$.  The value of $\rho_{H}(\vec G)$ in Eq.~(\ref{eq: rho components}) depends on the extent of the wavefunctions in momentum space, hence we can write the Fourier expansion of the Hartree potential in real space as
\begin{equation}
V_\text{H}(\vec{r})=V_{0}\sum_{n}\rho_\text{H}(\vec G_{n})e^{i\vec G_{n}\cdot\vec{r}}
\label{eq: VhRealSpace}
\end{equation}
with $\rho_\text{H}(\vec G_{n})=\left|\rho_\text{H}(\vec G_{n})\right|e^{i\arg\left[\rho_\text{H}(\vec G_{n})\right]}$ a complex number with $-\pi<\arg\left[\rho_\text{H}(\vec G_{n})\right]<\pi$ and $V_{0}=e^{2}/\varepsilon L$  the effective Coulomb potential. As shown in Ref.~\citep{Guinea2018a}, the dominant contribution to the Hartree potential comes from the first star, 
$\boldsymbol{G}_{n}=\pm \boldsymbol{G}_{1},\pm \boldsymbol{G}_{2},\pm (\boldsymbol{G}_{1}+\boldsymbol{G}_{2})$, of reciprocal lattice vectors. In addition, by exploiting the $C_6$ symmetry, the Fourier components of the Hartree potential are equally weighted. To solve the self-consistent Hartree Hamiltonian, the charge distribution is approximated as $\rho_{H}=\overline{\rho}_{H}+\delta\rho_{H}$ where $\overline{\rho}_{H}$ is a constant which takes into account the total density from all bands not included in the calculations~\citep{Guinea2018a}. The charge distribution is fixed by considering an homogeneous state at the charge neutrality point, this is $\rho_{H}=0$. Therefore, the  integral in Eq.~(\ref{eq: rho components}) is evaluated only over energy levels with $E_l\left(\vec k\right)$ between the charge neutrality point and Fermi level. The matrix elements of the Hartree potential $V_H$ in Eq. (\ref{eq: rho components}) depends implicitly on the filling fraction $\nu$ of the conduction band. For  fully-filled valence and conduction bands we have $\nu=4$ and when they are both empty we have $\nu=-4$. 
For a given value of the filling, we calculate the miniband spectrum and wavefunctions with self-consistent diagonalization of the main Hamiltonian by considering a coupling up to 5 stars, this corresponds to a total of $N=91$ vectors in the reciprocal space.

\subsection{The Kohn-Luttinger formalism for TTG}\label{APP:KL}

\begin{figure}
\centering
\includegraphics[width=3.5in]{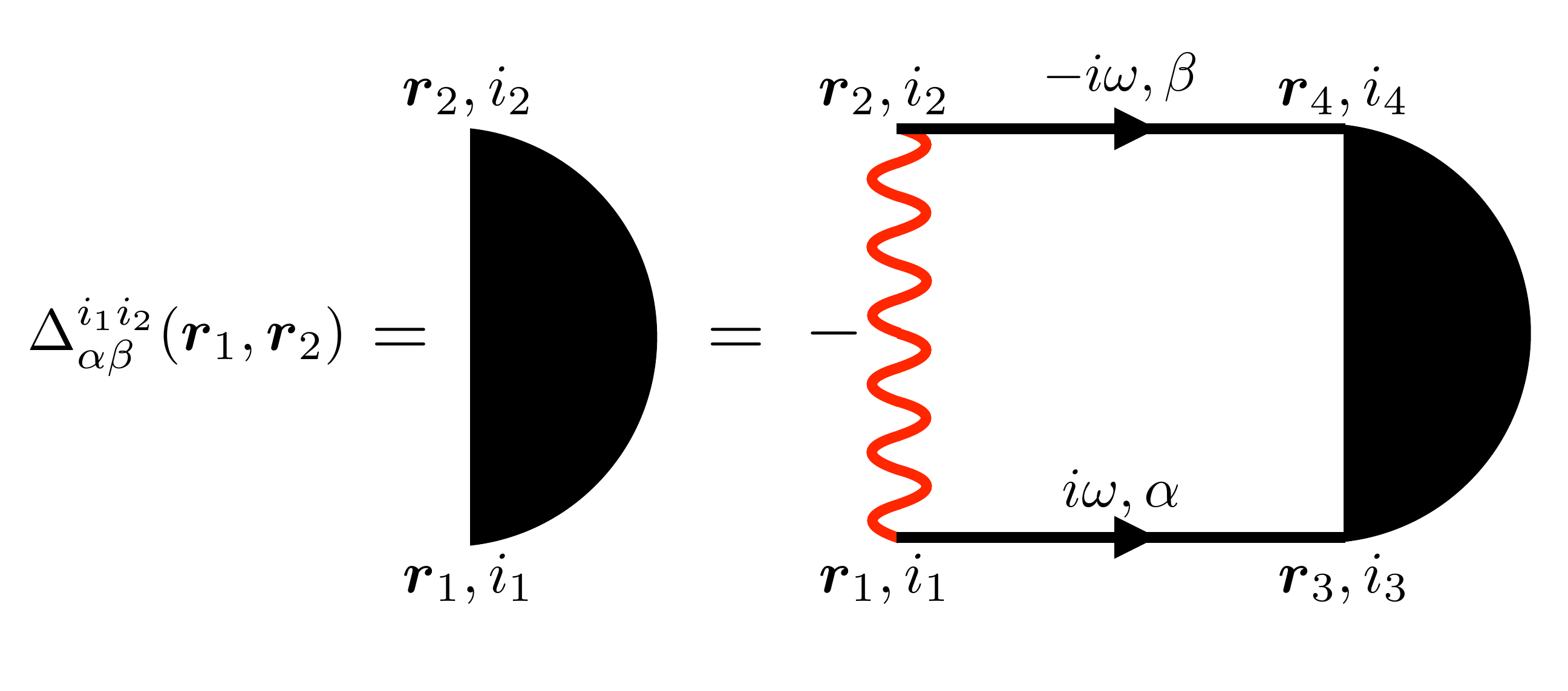}
\caption{
\textbf{Diagrammatic representation of the equation \pref{BCS_vertex_APP}}.
The wavy line denotes the screened potential and
the straight lines are the Green's functions.
}
\label{BCS_vertex_fig}
\end{figure}

Because of the lack of translational invariance, it is convenient to write the linearized equation for the order parameter (OP), $\Delta$, in real space:
\bea\label{BCS_vertex_APP}
\Delta^{i_1i_2}_{\alpha\beta}(\vec{r}_1,\vec{r}_2)=
-{\cal V}^{scr}(\vec{r}_1,\vec{r}_2)\int_\Omega\,d^2\vec{r}_3d^2\vec{r}_4
\sum_{i_3 i_4}K_BT\sum_{\omega}
\mathcal{G}^{i_1 i_3}_{\vec{r}_1 \vec{r}_3,\alpha}\left(i\hbar\omega\right)
\mathcal{G}^{ i_2 i_4}_{\vec{r}_2 \vec{r}_4,\beta}\left(-i\hbar\omega\right)
\Delta^{i_3 i_4}_{\alpha\beta}(\vec{r}_3 ,\vec{r}_4),
\eea
where
the $i$'s denote the layer and sublattice,
$\alpha\ne\beta$ is the index of flavor,
which encodes both the valley and spin degrees of freedom:
$\alpha=(\nu,\sigma)$,
$\Omega$ is the area of the system,
${\cal V}$ is the screened Coulomb potential,
accounting for the RPA corrections in both the particle-hole and the electron-phonon channels,
as detailed in the App. \ref{App:Vscr} below,
$T$ is the temperature,
$\omega$ are fermionic Matsubara frequencies and
$\mathcal{G}$ is the electronic Green's function calculated in the normal state.
The Eq. \pref{BCS_vertex_APP} is represented diagrammatically in the Fig. \ref{BCS_vertex_fig},
where the wavy line denotes the screened potential and
the straight lines are the Green's functions.

The Green's functions can be generally written as:
\bea
\mathcal{G}^{ij}_{\vec{r}\vec{r}',\alpha}(i\hbar\omega)=\sum_{n\vec{k}}\frac{\Phi^i_{n\vec{k},\alpha}(\vec{r})\Phi^{j,*}_{n\vec{k},\alpha}(\vec{r}')}{i\hbar\omega+\mu-E_{n\vec{k},\alpha}},
\eea
where $\vec{k}$ is the wave vector in the BZ, $\mu$ is the chemical potential and $\Phi^i_{n\vec{k},\alpha}$ is the Bloch eigenfunction corresponding to the band $E_{n\vec{k},\alpha}$:
\bea\label{Bloch_waves}
\Phi^i_{n\vec{k},\alpha}(\vec{r})=\frac{e^{i\vec{k}\cdot\vec{r}}}{\sqrt{\Omega}}
\sum_{\vec{G}}\phi^{i}_{n\vec{k},\alpha}(\vec{G})e^{i\vec{G}\cdot\vec{r}},
\eea
where
$\vec{G}=n_1\vec{G}_1+n_2\vec{G}_2$ are reciprocal lattice vectors, with $n_1,n_2$ integers,
and $\phi^{i}_{n\vec{k},\alpha}$ are numerical eigenvectors,
normalized according to: $\sum_{\vec{G}i}\phi^{i,*}_{n\vec{k},\alpha}(\vec{G})\phi^{i}_{m\vec{k},\alpha}(\vec{G})=\delta_{nm}$.
Note that the Green's function does not depend explicitly on the spin index,
as the continuum Hamiltonian, Eq. \pref{eq: Hamiltonian}, does not.

By performing the sum over the Matsubara frequencies in the Eq. \pref{BCS_vertex_APP}, one obtains:
\bea\label{BCS_vertex2}
\Delta^{i_1i_2}_{\alpha\beta}(\vec{r}_1,\vec{r}_2)&=&
-{\cal V}^{scr}(\vec{r}_1,\vec{r}_2)
\sum_{n_1\vec{k}_1 n_2\vec{k}_2}
\Phi^{i_1}_{n_1\vec{k}_1,\alpha}(\vec{r}_1)
\Phi^{i_2}_{n_2\vec{k}_2,\beta}(\vec{r}_2)
\left[
\frac{
f\left(-E_{n_2\vec{k}_2,\beta}+\mu\right)-f\left(E_{n_1\vec{k}_1,\alpha}-\mu\right)
}{E_{n_1\vec{k}_1,\alpha}+E_{n_2\vec{k}_2,\beta}-2\mu}\right]
\times\\
&\times&
\int_\Omega\,d^2\vec{r}_3d^2\vec{r}_4
\sum_{i_3 i_4}
\Phi^{i_3,*}_{n_1\vec{k}_1,\alpha}(\vec{r}_3)
\Phi^{i_4,*}_{n_2\vec{k}_2,\beta}(\vec{r}_4)
\Delta^{i_3i_4}_{\alpha\beta}(\vec{r}_3,\vec{r}_4),\nn
\eea
where: $f(\xi)\equiv \left(1+e^{\xi/K_BT}\right)^{-1}$ is the Fermi distribution.
Although the translational invariance is not preserved in general,
the screened potential, ${\cal V}^{scr}$, is still translationally invariant at the moir\'e scale,
which means:
${\cal V}^{scr}\left(\vec{r}_1,\vec{r}_2\right)={\cal V}^{scr}\left(\vec{r}_1+\vec{R},\vec{r}_2+\vec{R}\right)$
for any Bravais vector, $\vec{R}$, of the moir\'e lattice.
As a consequence, ${\cal V}^{scr}$ can be generally expressed in the Fourier basis as:
\bea
{\cal V}^{scr}\left(\vec{r}_1,\vec{r}_2\right)=\frac{1}{\Omega}\sum_{\vec{q}\vec{G}_1\vec{G}_2}
{\cal V}^{scr}_{\vec{G}_1,\vec{G}_2}\left(\vec{q}\right)
e^{i\vec{q}\cdot(\vec{r}_1-\vec{r}_2)}e^{i\vec{G}_1\cdot\vec{r}_1}e^{-i\vec{G}_2\cdot\vec{r}_2},
\eea
where $\vec{q}$ belongs to the moir\'e BZ and:
${\cal V}^{scr}_{\vec{G}_1,\vec{G}_2}\left(\vec{q}+\vec{G}\right)={\cal V}^{scr}_{\vec{G}_1+\vec{G},\vec{G}_2+\vec{G}}\left(\vec{q}\right)$.
Note that the off-diagonal elements, $\vec{G}_1\ne\vec{G}_2$, are triggered by the Umklapp processes,
whereas ${\cal V}^{scr}_{\vec{G}_1,\vec{G}_2}\left(\vec{q}\right)\sim\delta_{\vec{G}_1,\vec{G}_2}$ in the translational invariant limit.
Without loss of generality, we can then assume a similar Fourier expansion for the OP:
\bea
\Delta^{i_3i_4}_{\alpha\beta}\left(\vec{r}_3,\vec{r}_4\right)=
\frac{1}{\Omega}\sum_{\vec{q}\vec{G}_3\vec{G}_4}
\Delta^{i_3i_4}_{\alpha\beta;\vec{G}_3,\vec{G}_4}\left(\vec{q}\right)
e^{i\vec{q}\cdot(\vec{r}_3-\vec{r}_4)}e^{i\vec{G}_3\cdot\vec{r}_3}e^{-i\vec{G}_4\cdot\vec{r}_4},
\eea
which implies that only the terms with $\vec{k}_2=-\vec{k}_1$ survive in the rhs of the Eq. \pref{BCS_vertex2}.
Defining:
\begin{subequations}
\bea
\Delta^{n_1n_2}_{\alpha\beta}(\vec{q})&\equiv&
\int_\Omega\,d^2\vec{r}_3d^2\vec{r}_4
\sum_{i_3 i_4}
\Phi^{i_3,*}_{n_1\vec{q},\alpha}(\vec{r}_3)
\Phi^{i_4,*}_{n_2-\vec{q},\beta}(\vec{r}_4)
\Delta^{i_3i_4}_{\alpha\beta}(\vec{r}_3,\vec{r}_4),\\
\tilde{\Delta}^{n_1n_2}_{\alpha\beta}(\vec{q})&\equiv&
\Delta^{n_1n_2}_{\alpha\beta}(\vec{q})\times
\sqrt{
\frac{
f\left(-E_{n_2-\vec{q},\beta}+\mu\right)-f\left(E_{n_1\vec{q},\alpha}-\mu\right)
}{E_{n_2-\vec{q},\beta}+E_{n_1\vec{q},\alpha}-2\mu}
},
\eea
\end{subequations}
we project the Eq. \pref{BCS_vertex2} on the Bloch's eigenstates,
which gives the following equation for $\tilde{\Delta}^{n_1n_2}_{\alpha\beta}(\vec{q})$:
\bea\label{BCS_vertex_qspace}
\tilde{\Delta}^{m_1m_2}_{\alpha\beta}(\vec{k})=
\sum_{n_1n_2}\sum_{\vec{q}}\Gamma^{m_1m_2}_{n_1n_2;\alpha\beta}(\vec{k},\vec{q})
\tilde{\Delta}^{n_1n_2}_{\alpha\beta}(\vec{q}),
\eea
with:
\bea\label{kernel}
\Gamma^{m_1m_2}_{n_1n_2;\alpha\beta}(\vec{k},\vec{q})&=&
-\frac{1}{\Omega}\sum_{\vec{G_1}\vec{G_1}'}\sum_{\vec{G_2}\vec{G_2}'}\sum_{i_1i_2}
{\cal V}^{scr}_{\vec{G}_1-\vec{G}_1',\vec{G}_2-\vec{G}_2'}\left(\vec{k}-\vec{q}\right)
\phi^{i_1,*}_{m_1\vec{k},\alpha}(\vec{G}_1)
\phi^{i_2,*}_{m_2-\vec{k},\beta}\left(\vec{G}_2'\right)
\phi^{i_1}_{n_1\vec{q},\alpha}\left(\vec{G}_1'\right)
\phi^{i_2}_{n_2-\vec{q},\beta}(\vec{G}_2)\times\nn\\
&\times&
\sqrt{
\frac{
f\left(-E_{m_2-\vec{k},\beta}+\mu\right)-f\left(E_{m_1\vec{k},\alpha}-\mu\right)
}{E_{m_2-\vec{k},\beta}+E_{m_1\vec{k},\alpha}-2\mu}}\times
\sqrt{
\frac{
f\left(-E_{n_2-\vec{q},\beta}+\mu\right)-f\left(E_{n_1\vec{q},\alpha}-\mu\right)
}{E_{n_2-\vec{q},\beta}+E_{n_1\vec{q},\alpha}-2\mu}}.
\eea
The condition for the onset of superconductivity is that the kernel $\Gamma$ has the eigenvalue 1.
This defines the critical temperature, $T_c$, as the one at which the largest eigenvalue of $\Gamma$ is equal to 1.

Because the eigenfunctions do not depend explicitly on the spin index,
we can identify two different kinds of OP,
depending if $\alpha$ and $\beta$ share the same or opposite valley indices.
These two OP's describe intra-valley or inter-valley superconductivity, respectively. In the present work we focus on the inter-valley case, as we checked that the intra-valley superconductivity is less robust.

In order to diagonalize the kernel $\Gamma$, we project the Eq. \pref{BCS_vertex_qspace} onto the two bands in the middle of the spectrum, that give the main contribution to superconductivity.
For each filling, we include into $\Gamma$ the Hartree corrections on top of the non-interacting band structure,
as detailed in the Refs.~\citep{Guinea2018a,Cea2019}.
\subsection{Computation of the screened potential}\label{App:Vscr}

Here we provide the details concerning the calculation of the screened potential in the reciprocal space:
${\cal V}^{scr}_{\vec{G}_1,\vec{G}_2}\left(\vec{q}\right)$.
First, we introduce the unscreened Coulomb potential:
\bea
{\cal V}_C(\vec{q})=2\pi e^2\frac{\tanh\left(d_g|\vec{q}|\right)}{\epsilon|\vec{q}|},
\eea
where $e$ is the electron charge, $d_g$ is the distance of the sample from a metallic gate and $\epsilon$ is the relative dielectric constant of the environment.
In this work we use: $\epsilon=10$ and $d_g=40$nm.
We consider the effect of the strain in TTG by means of longitudinal acoustic phonons coupling to the charge density via the deformation potential:
$v_d(\vec{r})=-D s(\vec{r})$, where $s(\vec{r})$ is the local strain and $D$ is the electron-phonon coupling, for which we use: $D=16$eV.
We compute the screened potential, including the RPA corrections induced by both the particle-hole excitations and the electron-phonon coupling.
In matrix notation, the expression for the inverse of ${\cal V}^{scr}$ is given by:
\bea\label{inverseVscr}
\left[{\cal V}^{scr}(\vec{q})\right]^{-1}=\left[\hat{{\cal V}}_C(\vec{q})\right]^{-1}-\chi^0(\vec{q})+
g\chi^0(\vec{q})\left[\mathbb{1}+g\chi^0(\vec{q})\right]^{-1}\chi^0(\vec{q}),
\eea
where the entries are indexed by the reciprocal lattice vectors,
$\left[\hat{{\cal V}}_C(\vec{q})\right]_{\vec{G}_1,\vec{G}_2}\equiv {\cal V}_C\left(\vec{q}+\vec{G}_1\right)\delta_{\vec{G}_1,\vec{G}_2}$,
$g=\frac{D^2}{\lambda+2\mu}$,
 with $\lambda=3.25$ eV\AA$^{-2}$, $\mu=9.44$ eV\AA$^{-2}$ the Lam\'e coefficients of monolayer graphene,
and $\chi^0(\vec{q})$ is the static density-density response function,
which is given by:
\bea
\chi^0_{\vec{G}_1,\vec{G}_2}\left(\vec{q}\right)&=&\frac{1}{\Omega}\sum_{\vec{k}\vec{G}'_1\vec{G}'_2}\sum_{n_1n_2}\sum_{ij\alpha}
\phi^i_{n_1\vec{k}+\vec{q},\alpha}\left(\vec{G}'_1+\vec{G}_1\right)\phi^{i,*}_{n_2\vec{k},\alpha}\left(\vec{G}'_1\right)
\phi^{j,*}_{n_1\vec{k}+\vec{q},\alpha}\left(\vec{G}'_2+\vec{G}_2\right)\phi^j_{n_2\vec{k},\alpha}\left(\vec{G}'_2\right)\times\\
&\times&
\frac{f\left( E_{n_2\vec{k},\alpha}-\mu \right)-f\left( E_{n_1\vec{k}+\vec{q},\alpha}-\mu \right)}
{ E_{n_2\vec{k},\alpha}-E_{n_1\vec{k}+\vec{q},\alpha}}.
\nn
\eea
To compute $\chi^0$ numerically, we include the 10 bands closest to the charge neutrality point.

\end{document}